\documentclass[a4paper,11pt]{article}
\pdfoutput=1

\usepackage{amsmath}
\usepackage{graphicx}
\usepackage{amsfonts}
\usepackage{amssymb}
\usepackage{xcolor}

\usepackage[colorlinks=true, urlcolor=violet, linkcolor=blue,
  citecolor=red, hyperindex=true, linktocpage=true]{hyperref}

\newcommand{\bc}{\begin{center}}
\newcommand{\ec}{\end{center}}
\def\ba#1{\begin{array}{#1}\displaystyle}
\newcommand{\ea}{\end{array}}
\newcommand{\z}{\\[2mm] \displaystyle}

\newcommand{\beq}{\begin{equation}}
\newcommand{\eeq}{\end{equation}}
\newcommand{\beqa}{\begin{eqnarray}}
\newcommand{\eeqa}{\end{eqnarray}}
\newcommand{\no}{\nonumber}
\newcommand{\n}{\nonumber\\}
\newcommand{\bi}{\begin{itemize}}
\newcommand{\ei}{\end{itemize}}

\def\lt#1{\left#1}
\def\rt#1{\right#1}
\def\t#1{\tilde{#1}}
\def\h#1{\hat{#1}}
\def\b#1{\bar{#1}}
\def\frc#1#2{\frac{#1}{#2}}

\newcommand{\p}{\partial}

\newcommand{\vac}{{\rm vac}}
\newcommand{\bra}{\langle}
\newcommand{\ket}{\rangle}
\newcommand{\Z}{{\mathbb{Z}}}

\newcommand{\R}{{\mathbb{R}}}

\newcommand{\Or}{{\cal O}}

\newcommand{\ep}{\epsilon}

\newcommand{\Tr}{{\rm Tr}}

\newcommand{\qq}{{\rm q}}

\newcommand{\rl}{{\rm L}}
\newcommand{\rr}{{\rm R}}

\begin{document}

\begin{titlepage}

\begin{center}
{\Large {\bf Non-equilibrium steady states in the Klein--Gordon theory}

\vspace{1cm}

Benjamin Doyon$^{1,*}$, Andrew Lucas$^2$, \\ 
Koenraad Schalm$^{2,3}$ and M. J. Bhaseen$^4$}

$^1$ Department of Mathematics, King's College London, Strand, London, U.K.\\
$^2$ Department of Physics, Harvard University, Cambridge, MA 02138 USA.\\
$^3$ Instituut-Lorentz, Leiden University, Leiden, The Netherlands.\\
$^4$ Department of Physics, King's College London, Strand, London, U.K.\\
$^*$ email: benjamin.doyon@kcl.ac.uk

\end{center}

\vspace{1cm}

\noindent We construct non-equilibrium steady states in the
Klein--Gordon theory in arbitrary space dimension $d$ following a
local quench. We consider the approach where two independently
thermalized semi-infinite systems, with temperatures $T_{\rm L}$ and
$T_{\rm R}$, are connected along a $d-1$-dimensional hypersurface. A current-carrying steady state, described by thermally distributed modes with temperatures $T_{\rm L}$ and $T_{\rm R}$ for left and right-moving modes, respectively, emerges at late times. The non-equilibrium density matrix
is the exponential of a non-local conserved charge. We obtain exact
results for the average energy current and the complete distribution
of energy current fluctuations.  The latter shows that the long-time
energy transfer can be described by a continuum of independent Poisson
processes, for which we provide the exact weights. We further describe
the full time evolution of local observables following the quench. Averages of generic local observables, including the stress-energy tensor, approach the
steady state with a power-law in time, where the exponent depends
on the initial conditions at the connection hypersurface.
We describe boundary conditions and special operators for
which the steady state is reached
  instantaneously on the connection hypersurface.
A semiclassical analysis of freely propagating modes yields the
average energy current at large distances and late times. We conclude by comparing and
contrasting our findings with results for interacting theories and
provide an estimate for the timescale governing the crossover to
hydrodynamics. As a
modification of our Klein-Gordon analysis we also include exact
results for free Dirac fermions.   \vfill

{\ }\hfill \today

\end{titlepage}

\tableofcontents

\section{Introduction}

Understanding far-from-equilibrium phenomena is one of the most
important challenges of current theoretical physics research. Amongst
these phenomena, non-equilibrium steady states (NESS), involving
constant flows of energy, particles or charge between leads, play an
important role in both theory and experiment. Although stationary,
they exhibit many of the non-trivial features associated with
non-equilibrium physics, including generalized fluctuation relations
\cite{gallavotti, jarzynski, esposito}. The study of quantum NESS is
of particular interest, especially in the presence of emergent
collective behavior, as it sheds light on the interplay between
quantum effects and non-equilibrium physics.

From a theoretical perspective, there are a variety of ways
to represent NESS.
Here, we consider the partitioning approach, or Hamiltonian-reservoir
formulation, where the baths are fully and exactly represented. This
is a real-time construction of NESS, whereby two infinitely long leads
are initially thermalized in different equilibrium states, 
and are then suddenly connected (either to another quantum system, or
just to each other) and allowed to evolve unitarily for a long time.
Such constructions have been used in a variety of different
contexts. For example, in combination with Keldysh perturbation
theory, they have been used in order to study charge currents through
nanostructures \cite{caro}. They have also been used to study thermal
flows in infinite classical chains of harmonic oscillators
\cite{rub,spo}. Quantum transport has also been investigated within
the $C^*$-algebra formalism for the free-fermionic \cite{tas, tas2}
and XY \cite{ah,og,aschp} quantum chains; for general results see
\cite{ruelle,cor}. In modern parlance within the physics community,
this formulation is a ``local quench'' connecting initial mixed
states.

The Hamiltonian-reservoir formulation is particularly well adapted to
the study of the interplay between quantum collective behavior and
non-equilibrium physics, as it is readily formulated within quantum
field theory (QFT). In the context of quantum impurity models, where
the leads are described by fermionic seas with free gapless
excitations, this point of view has been very successful. Various
aspects of perturbation theory
\cite{ram,meir,hers,jauho,doyand,doyirlm,freboul}, integrability
\cite{fen,fen2,boul,kom,car} and non-equilibrium fluctuation theory
\cite{lev,kl,scho,avr,kom,car,berjmp} have been developed.  More
recently, attention has focused on models where the leads exhibit
non-trivial emergent properties. Non-equilibrium flows, including the
full fluctuation theory, have been studied in homogeneous critical
systems described by conformal field theory (CFT)
\cite{gut,gut2,doyon1d1, doyon1d2}, in universal regions described by
integrable QFT \cite{doyonint,doyqft}, and in the quantum Ising model
\cite{doyonising}.  The CFT predictions have been confirmed
numerically using time-dependent density matrix renormalization group
simulations of quantum spin chains \cite{Moore1, joelmoore}.

In all of these examples, the
system considered is one-dimensional. The generalization to
$d$-dimensional leads connected along a $d-1$ dimensional hypersurface
is quite non-trivial due to the absence of widely applicable exact
techniques. Recent progress on this problem was achieved for strongly
interacting quantum critical models \cite{bhaseen1} using a
combination of insights derived from QFT, gauge-gravity duality
and fluid dynamics. It was
shown that the NESS for thermal transport in these interacting quantum 
systems is fully described by a Lorentz boosted thermal state. 

In this work, we focus on the opposite limit of vanishing
interactions, by considering the 
free massive Klein--Gordon theory 
in $d$-dimensions, with $d\geq 1$. The complete
time-evolution, including the emergence of a NESS, is fully amenable
to theoretical treatment.  This model
describes the low-energy scaling limit of an array of coupled
harmonic oscillators, and as such provides a paradigmatic model for
studying thermal transport in detail.  After the original works
\cite{rll,spo,na} in the classical realm, quantum harmonic crystals
have been studied within various formalisms
\cite{zt,saito,snh,dr,sd,rd,dsh}. This also includes studies in higher
dimensions \cite{na,na2,rd}, with the restriction that the transverse
directions be finite. Here, we consider the 
far from equilibrium response of the genuinely infinite 
continuum CFT with Lorentz invariance.

Whilst this work was in preparation,
results in higher-dimensional non-relativistic and massless
relativistic free fermion models have been obtained \cite{collura2},
where efficient semi-classical-type techniques are developed for
averages of single local observables. Here we provide instead a full
quantum calculation; this allows us to go beyond averages of single
observables and to obtain a rather complete description of the
Klein--Gordon model. This includes a proof and analysis of the
non-equilibrium density matrix, the full current fluctuation spectrum
and the approach to the steady state. We supplement our analysis with
new exact results for free Dirac fermions.

A key motivation for the present work is to compare and contrast the
free-field limit with our previous results in the strongly interacting
regime \cite{bhaseen1}. We will show that in higher dimensions, the
critical Klein-Gordon model does {\em not} reproduce the strongly
interacting results of \cite{bhaseen1}, and nor should it: there is a
fundamental distinction between the behavior of free-fields and
interacting CFT in $d>1$. This is different from the one-dimensional
case, where the free-field limit captures the general CFT results.  A
crucial distinction in higher-dimensions is that the Klein--Gordon
model contains infinitely many conserved quantities, in stark contrast
to a generic higher-dimensional CFT;
this has a direct impact on the non-equilibrium density matrix and
thus physical observables. Indeed, it is well known that harmonic
crystals exhibit anomalous non-equilibrium transport properties due to
this proliferation of conservation laws, including the inapplicability
of Fourier's law.
The recent results of \cite{doyon1d1,doyon1d2,bhaseen1} show that
anomalous transport also occurs in interacting quantum critical
systems due to the presence of ballistic transport. 
The present paper indicates that in higher dimensions,
quantum critical systems
display
more acutely
anomalous behavior: there is a disconnect between the
free-field limit and the generic 
interacting problem. We shall discuss the
role of interactions in bridging these results. Our exact free-field
results may also provide useful benchmarks for numerical simulations
in more than one dimension.

The specific situation we shall consider is one where the two
semi-infinite halves of the model (i.e. the ``leads''), at $x^1<0$ and
$x^1>0$, are independently thermalized at temperatures $T_{\rm L}$ and
$T_{\rm R}$, respectively~\cite{doyon1d1,bhaseen1}.  They are then
brought into instantaneous contact along the $d-1$ dimensional
hypersurface $x^1=0$ and are allowed to evolve unitarily; see
Fig.~\ref{Fig:Quench}.  We consider the impact of different initial
conditions at the connection hypersurface, including both free and
fixed. We find the following results, some of which were also found
(in different forms) in \cite{collura2}:
\begin{figure}
\begin{center}
\includegraphics[width=14cm]{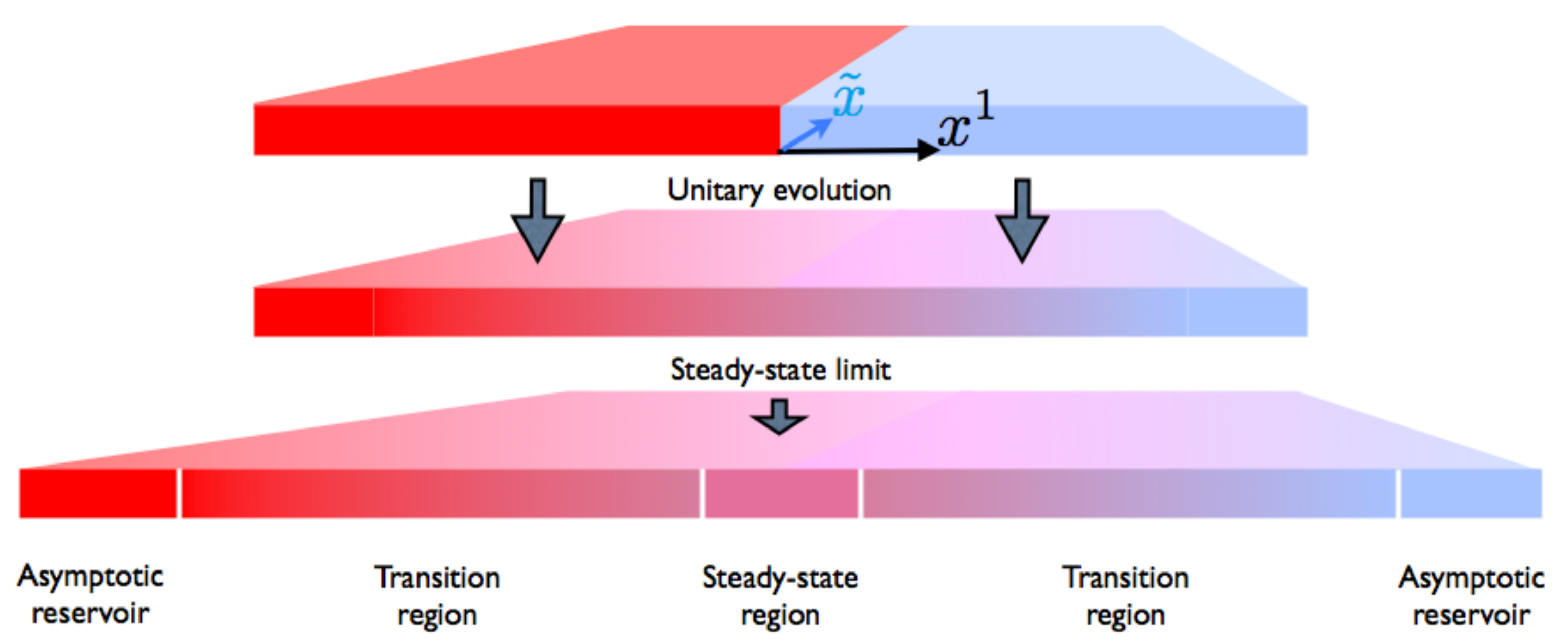}
\caption{Two semi-infinite systems described by the massive
  Klein--Gordon model in $d$ dimensions are independently thermalized
  at temperatures $T_{\rm L}$ and $T_{\rm R}$. They are brought into
  instantaneous contact along the $d-1$ dimensional hypersurface $x^1=0$. At
  late times a spatially homogeneous non-equilibrium steady state
  (NESS) emerges carrying a non-vanishing energy current.}
\label{Fig:Quench}
\end{center}
\end{figure}
\bi
\item At late times, 
a steady state carrying an energy
  current emerges. It is described by thermally distributed modes of
  positive and negative longitudinal momenta with temperatures $T_{\rm
    L}$ and $T_{\rm R}$ respectively. The form of the steady
    state is independent of the initial conditions on the
  hypersurface and the exact energy current separates into a
  difference of a function 
evaluated at
  $T_{\rm L}$ and $T_{\rm R}$ respectively. Similar
  observations were made in other free-particle models in one
  dimension \cite{tas, tas2, ah,og,aschp,doyonising} and in higher dimensions
  \cite{collura2}; this separation is in contrast to the
    behavior expected for an interacting CFT in $d>1$ \cite{bhaseen1}.
\item One may describe the exact steady state using a density
  matrix. The density matrix has the form $\rho_{\rm s} = \mathrm{e}^{-\h W}$
  where $\h W$ is a conserved quantity. This conserved quantity is
  {\em not} local: it is rather bilinear in the fields, with an
  algebraically decaying kernel. This points to {\em asymptotically
    algebraic correlations} for certain observables 
in the non-equilibrium steady state.
\item All the cumulants of the total energy transferred, within
  a time $t$ and through a transverse hypersurface of area $A$, scale
  like $t A$ as $t,A\to\infty$. The scaled cumulants are obtained by
  dividing by $t A$ and taking the limit $t,A\to\infty$.  The
  exact generating function for all the scaled cumulants is that of a
  continuum of Poisson processes with a weight function $\omega(q)$ that
  is analytically determined; see Eq.~\eqref{omega}.
That is, the scaled cumulants are given by $\int \mathrm{d}q\,\omega(q) \,q^k$,
including the average energy current at $k=1$.
\item The averages of generic local observables approach the NESS with
  a power-law in time $t$, where the exponent depends on the initial
  conditions at the connection hypersurface. The generic leading
    order response is faster, $t^{-1}$, for free initial boundary
  conditions, and slower, $t^{-\frc12}$, for other initial
  conditions. The vanishing of the leading order contribution may give
  rise to a faster approach.

\item The energy current and energy density approach their steady-state
  values faster than the generic result in the case of fixed boundary
  conditions: $t^{-2}$ for the average of the energy current, and
  $t^{-1}$ for the average of the energy density, with oscillating
  factors. They agree with the generic result for free boundary
  conditions.

\item For free initial boundary conditions, the steady state is
  reached {\em immediately} for any observable lying {\em on the
    connection hypersurface itself} and hyper-local (without derivatives) in the longitudinal direction. At the level of averages of local
  fields, this is similar to the results found in \cite{collura2}. For
  fixed initial boundary conditions a similar statement holds for observables involving only single derivatives of fundamental fields in the longitudinal direction, and these statements generalize to observables involving only even or only odd derivatives of fundamental fields.

\item In the massless case, fields integrated along the transverse
  direction take their steady state values immediately inside the
  light-wedge. This can be explained by a suitable dimensional
  reduction of the Klein--Gordon theory and results for $d=1$ CFT. A similar dimensional reduction was at the basis of the methods used in \cite{collura2}; in Appendix \ref{app1d} we provide a full operator statement  
of this dimensional reduction within the Klein-Gordon model.
\item At space-time positions far from the connection event and
  within the light-wedge, a semiclassical analysis can be used, where
  averages receive independent contributions from waves traveling at
  their respective group velocities, carrying the thermal information
  from the left and right reservoirs. In this region, averages of
  local observables take simple forms, and in particular are monotonic
  with time. A similar semiclassical picture is used in
  \cite{collura2}. \ei

The layout of this paper is as follows. In Section \ref{Sect:Steady}
we describe the NESS in arbitrary dimensions, including exact results
for the average energy current, the average energy density, and the
entire spectrum of fluctuations. We interpret the results in terms of
a Poisson process. We also analyze the locality properties of the
non-equilibrium density matrix. In Section \ref{Sect:Time} we obtain
exact results for the complete time-evolution and extract the
power-law approach to the NESS, including the effects of the initial
boundary conditions.  We also consider the time evolution of
non-generic observables, including analytical results based on a
suitable dimensional reduction of the Klein--Gordon model.  We
complement our exact calculations with semiclassical results that are
valid in the limit of large distances and long times.  We conclude in
Section \ref{Sect:Conc}, presenting arguments for how the presence of
weak interactions will modify the free-field results 
engendering a crossover to 
the hydrodynamic regime \cite{bhaseen1}. We 
  include supporting calculations in Appendices
  \ref{appcal}-\ref{appsemi}, and also present the free Dirac fermion
  results in Appendix \ref{appfermion}.

\section{Exact steady-state density matrix, averages and fluctuations}
\label{Sect:Steady}

In this section we derive the exact average energy current and energy
density, as well as the full fluctuation spectrum, in the steady state
that occurs as the final configuration after the two semi-infinite
systems have been brought into contact. We employ the exact steady
state density matrix which we initially motivate on physical grounds
and derive later in Section \ref{Sect:Time}.

First, let us discuss the occurrence of a non-trivial steady state,
the notion of a steady state density matrix, and its expected form in
free models.

As is suggested by Fig.~\ref{Fig:Quench}, one expects that, due to
causality and finite propagation speeds, there will be at all times
regions which have not been affected by the quench and are still
thermal. A long time after the quench, these effective reservoirs are
far from the connection hypersurface. Hence, one would expect very
small energy density gradients in the region around the connection
hypersurface, and zero gradients at infinite time. This implies that
any diffusive energy transport will be suppressed, so that only
ballistic transport will occur in the steady state. Ballistic
transport is expected to occur, for instance, when the current is a
conserved density, due to Mazur's inequality \cite{maz,zot}, an
indicator of near-equilibrium ballistic transport. This is the case
for the energy current in any relativistic quantum field theory as it
is the density of the momentum, a conserved quantity. Hence, in this
case we expect to have a non-trivial current carrying steady state in
the partitioning approach. This argument holds in any dimension, and
the emergence of a non-zero current in field theory has been explicitly shown in one-dimensional CFT \cite{doyon1d1,doyon1d2},
in higher-dimensional CFT \cite{bhaseen1},
and in higher-dimensional free
fermion models \cite{collura2}. Section \ref{Sect:Time} provides a
proof in the higher-dimensional Klein-Gordon theory. Ballistic energy
transport is also expected in integrable lattice models that are not
necessarily in the scaling limit described by QFT
\cite{zot}. This has been explicitly shown only in
  one-dimensional models admitting free-boson or free-fermion
  representations, such as the harmonic chain \cite{rll,spo,zt},
  free-fermionic conductors \cite{tas, tas2} and the XY model
  \cite{ah,og,aschp}.

The steady state occurs, after an infinite time, in a region around
the connection hypersurface that is small as compared to the size of
the system; see Fig. \ref{Fig:Quench}. The steady state is a map
\[
	\Or \mapsto \bra\Or\ket_{\rm s}
\]
from observables lying in the steady state region, to their averages,
calculated in the steady-state limit; see Eq.~\eqref{limit}
below. This map does not provide any precise information about the
rest of the system. The steady-state region can be taken as any volume
of space extending a finite distance $\ell$ from the hypersurface
(here we concentrate solely on bulk observables, with the
boundaries of the system being asymptotically far from the
observables), and in the steady state limit we take $\ell\ll t \ll L$,
where $t$ is the time after the connection, and $L$ the linear size of
the system.

In the limit $L\to\infty$ followed by $t\to\infty$, although the
region is small as compared to the system, one can take $\ell$ as
large as possible. In fact, in order to describe the full steady state
and analyze, for instance, the large-distance behavior of correlation
functions, one must take $\ell\to\infty$. The steady state then lies
on an infinite (open) system.

A standard way of describing quantum averages is by using a density
matrix and tracing over a Hilbert space ${\cal H}$: \beq\label{tror}
\bra\Or\ket_{\rm s} = \frc{\Tr_{{\cal H}}\lt(\rho_{\rm s}
  \Or\rt)}{\Tr_{{\cal H}}\lt(\rho_{\rm s}\rt)}.  \eeq Intuitively, one
may extract two ingredients in this description: the density matrix
$\rho_{\rm s}$ and the tracing operation $\Tr_{\cal H}$. Physically,
one may expect the former to contain the information of the state the
system is in, and the latter that of the dynamics of the
system. Steady states of open quantum systems present problems: in
non-equilibrium steady states, the volume is intrinsically infinite,
and there are no {\em a priori} finite-volume underlying descriptions
giving discrete sets of vectors in which to define the trace operation
$\Tr_{\cal H}$ and the density matrix $\rho_{\rm s}$.

Nevertheless, as advocated in \cite{ruelle}, one may use a continuum
of (un-normalizable) scattering states, instead of a discrete
basis. In an IR-free QFT, these are the asymptotic states,
representing massive relativistic particles that are asymptotically
freely propagating. In the case of energy transfer, the eigenvalues of
the steady state density matrix $\rho_{\rm s}$ on the asymptotic
states were proposed in \cite{doyon1d1,doyqft} in one-dimensional
models. They describe weights according to the total energy of
particles with positive and negative momenta. Weights for particles of
positive (negative) momenta are Boltzmann weights at temperature
$T_{\rm L}$ ($T_{\rm R}$). This description is a generalization of
what was found in the free-fermionic and XY chains
\cite{tas,ah,og,aschp}.

This description is, in general, not enough to fully determine the map
$\Or\mapsto \bra\Or\ket_{\rm s}$ associated with the density matrix as
per \eqref{tror}: because of the continuum of asymptotic states, one
needs to determine their density in order to define the trace
operation $\Tr_{\cal H}$, carrying the dynamical
information. This is in general a hard problem. A solution was
proposed for 1+1-dimensional integrable massive QFT in \cite{doyonint}
via (a generalization of) the thermodynamic Bethe ansatz, and the
problem is solved thanks to chiral factorization in one-dimensional
CFT \cite{doyon1d1, doyon1d2}.  The answer is quite different in
higher-dimensional interacting conformal field theory, where the long
time behavior is not characterized by a quasiparticle
description: it was found recently in \cite{bhaseen1} that the
emergent steady state must be a Lorentz boosted thermal state, with
the rest-frame temperature and boost velocity 
determined by gauge-gravity duality and 
relativistic hydrodynamics.

Here we consider the Klein-Gordon theory of a free boson.  Due to the
lack of interaction between particles, one may determine the density
of asymptotic states by thermalization of the independent Fock
modes. In Section \ref{Sect:Time} we explicitly show that this steady
state, in the $d$-dimensional Klein-Gordon theory, is indeed that
resulting from the partitioning approach, paralleling the situation in
one-dimensional quadratic and free-particle models. Below we define
and use this density matrix in order to obtain steady-state averages
of the energy current and energy density, as well as all the long-time
fluctuations of the energy transfer. We then analyze the
non-local form of the density matrix.

\subsection{The exact steady-state density matrix}\label{ssectdens}

In order to fix notations, we recall that the massive
Klein--Gordon model describes canonical fields $\phi(x)$ and $\pi(x)$
(for $x\in \R^d$ the space coordinate) with equal-time commutation
relations \beq\label{cr} [\phi(x),\phi(y)] = [\pi(x),\pi(y)]=0,\quad
[\phi(x),\pi(y)] = \mathrm{i}\delta^d(x-y) \eeq and Hamiltonian \beq\label{H} H
= \frc12 \int \mathrm{d}^dx\,:\big(\pi(x)^2+(\nabla\phi(x))^2 +m^2
\phi(x)^2\big):\;. \eeq We introduce the Fourier modes $A_p$, $A^\dag_p$
through \beq \phi(x)= \int D p \,(A_p \mathrm{e}^{\mathrm{i}p\cdot x} + A^\dag_p
\mathrm{e}^{-\mathrm{i}p\cdot x}),\quad \pi(x)= -\mathrm{i}\,\int D p \,E_p\, (A_p \mathrm{e}^{\mathrm{i}p\cdot x}
- A^\dag_p \mathrm{e}^{-\mathrm{i}p\cdot x}),
	\label{modeexp}
\eeq where $p\cdot x = \sum_{j=1}^d p^j x^j$, $E_p = \sqrt{p^2+m^2}$ is the relativistic energy for a
particle of momentum $p$, and $Dp = \frc{\mathrm{d}^d p}{(2\pi)^d\,2E_p}$ is
the Lorentz invariant measure.  The canonical commutation relations
\eqref{cr} imply \beq\label{crm} [A_p,A_q] = 0,\quad [A_p,A^\dag_q] =
(2\pi)^d\,2E_p\,\delta^d(p-q) \eeq and the Hamiltonian is diagonalized
as $H =\int Dp\,E_p\,A^\dag_p A_p$.  The associated Hilbert space is the Fock
space over the algebra \eqref{crm}, with vacuum $|\vac\ket$ defined by
$A_p|\vac\ket =0$.

The total Hamiltonians for the separate left and right subsystems can
be expressed as \beq\label{HLR} H_{\rm L,R} = \frc12 \int_{x_1\lessgtr
  0} \mathrm{d}^dx \,:\big(\pi(x)^2+(\nabla\phi(x))^2 + m^2\phi(x)^2\big):\;.
\eeq The initial density matrix, where the left and right subsystems
are independently thermalized at inverse temperatures $\beta_{\rl}$
and $\beta_{\rr}$ respectively, is then \beq\label{rho0} \rho_0 =
\mathrm{e}^{-\beta_{\rm L} H_{\rm L} - \beta_{\rm R} H_{\rm R}}.  \eeq In the
quench setup, we instantaneously connect the two halves together and
let the full system evolve unitarily with $e^{iHt}$ for a long period
of time. Given a product of local observables $\Or$, whose support is
finite, its average in the steady state is defined by the following
limit, if it exists: \beq\label{limit} \bra\Or\ket_{\rm s} =
\lim_{t\to\infty} \Tr\lt( \frak{n}[\rho_0] \,\mathrm{e}^{\mathrm{i}Ht}\Or \mathrm{e}^{-\mathrm{i}Ht}\rt),
\eeq where we use $\frak{n}[\rho]=\rho/\Tr(\rho)$. Below we also use the notation $\bra \Or\ket = \Tr\lt(\frak{n}[\rho_0]\,\Or\rt)$ for averages in the initial state, and operators are implicitly evolved with the dynamics generated by $H$: $\Or(t) = \mathrm{e}^{\mathrm{i}Ht}\Or \mathrm{e}^{-\mathrm{i}Ht}$.

According to the above discussion, we expect that the limit
\eqref{limit} exists for any product of local observables $\Or$, and
that its result can be calculated using the steady-state density
matrix \beq\label{rhoness} \rho_{\rm s} := \exp\lt[-\beta_\rl
  \int_{p^1>0} Dp\,E_p \,A^\dag_p A_p - \beta_\rr \int_{p^1<0} Dp\,E_p
  \,A^\dag_p A_p\rt].  \eeq This describes thermally distributed modes
of positive and negative longitudinal momenta with temperatures
$T_{\rm L}$ and $T_{\rm R}$ respectively. The steady-state 
density matrix is both stationary and homogeneous. Expectation values in
the NESS are given by \beq\label{ness} \bra\Or\ket_{\rm s} = \Tr
\lt(\frak{n}[\rho_{\rm s}] \Or\rt).  \eeq These may be evaluated using
the mode expansion \eqref{modeexp}, Wick's theorem and the
contractions \beq \Tr \lt(\frak{n}[\rho_{\rm s}] A_p A^\dag_q\rt) =
\frc{(2\pi)^d\,2E_p\,\delta^d(p-q)}{1-\mathrm{e}^{-W(p)}},\quad \Tr
\lt(\frak{n}[\rho_{\rm s}] A_p^\dag A_q\rt) =
\frc{(2\pi)^d\,2E_p\,\delta^d(p-q)}{\mathrm{e}^{W(p)}-1}.
	\label{traa}
\eeq
The latter are obtained using the cyclic property of the trace, the canonical commutation relations \eqref{crm}, and the exchange relation $\rho_{\rm s}A_p  = \mathrm{e}^{-W(p)}\,A_p \rho_{\rm s}$, where 
\[
	W(p):= \lt\{\ba{ll} \beta_{\rm L} E_p & (p^1>0) \\ \beta_{\rm R} E_p & (p^1<0).\ea\rt.
\]
In these notations
\[
	\rho_{\rm s} = \exp \lt[-\int Dp\,W(p)A_p^\dag A_p\rt].
\]

In Section \ref{Sect:Time} we will explicitly show that the steady
state density matrix $\rho_{\mathrm{s}}$ arises at late times in the quench
problem depicted in Fig.~\ref{Fig:Quench}.  In addition, we will
provide the complete time evolution of general averages of local
observables. In the remainder of this section we focus on the
consquences of the exact steady state density matrix, providing results for
the average energy density, energy current and the complete
distribution of energy current fluctuations.

\subsection{Steady-state averages of the energy current and energy density}

From the above description, the averages of the energy current density
$T^{01}$ and the energy density $T^{00}$ are readily evaluated.  The
stress-energy tensor is given by \beq\label{tmunu} T^{\mu\nu} =
:\p^\mu \phi \,\p^\nu \phi - \frc12 \eta^{\mu\nu} \, \p^\rho\phi
\,\p_\rho\phi:\;.  \eeq Using equations \eqref{modeexp}, \eqref{ness} and
\eqref{traa}, the result is \beq \bra T^{\mu\nu}\ket_{\rm s} =\int
Dp\, \frc{2p^\mu p^\nu}{\mathrm{e}^{W(p)}-1}, \eeq where $p^0 = E_p$. In
particular, the energy current density is that expected for a free bosonic
model, averaging the momentum with bosonic filling fractions dependent
on its sign: \beq\label{T01} \bra T^{01}\ket_{\rm s} = \int \frc{\mathrm{d}^d
  p}{(2\pi)^d} |p^1| \lt( \frc{\Theta(p^1)}{\mathrm{e}^{\beta_\rl E_p}-1} -
\frc{\Theta(-p^1)}{\mathrm{e}^{\beta_\rr E_p}-1}\rt).  \eeq

One may evaluate the averages of the energy current and the energy
density explicitly by performing the angular integrals; see Appendix
\ref{appcal}. The results may be expressed in various ways: \beqa \bra
T^{01}\ket_{\rm s} &=&
\frc{\Gamma\lt(\frc{d}2\rt)}{4\pi^{d/2+1}(d-1)!}  \int_0^\infty
\mathrm{d}p\,p^d \,\frc{\sinh\lt(\frc{\beta_{\rm R}-\beta_{\rm L}}2
  E_p\rt)}{\sinh\lt(\frc{\beta_{\rm
      R}}2E_p\rt)\sinh\lt(\frc{\beta_{\rm L}}2E_p\rt)}\n &=&
\frc{d\,\Gamma\lt(\frc d2\rt)}{2\pi^{d/2+1}}\,\lt( \zeta_{\frc m
  {T_\rl}}(d+1)\,T_\rl^{d+1}- \zeta_{\frc m
  {T_\rr}}(d+1)\,T_\rr^{d+1}\rt),\n
  \bra T^{00}\ket_{\rm s} &=&
\frc{\Gamma\lt(\frc{d+1}2\rt)}{4\pi^{(d+1)/2}(d-1)!}  \int_0^\infty
\mathrm{d}p\,p^{d-1}E_p \,\frc{\cosh\lt(\frc{\beta_{\rm R}-\beta_{\rm L}}2
  E_p\rt)}{\sinh\lt(\frc{\beta_{\rm
      R}}2E_p\rt)\sinh\lt(\frc{\beta_{\rm L}}2E_p\rt)}\n &=& \frc{d
  \,\Gamma\lt(\frc{d+1}2\rt)} {2\pi^{(d+1)/2}}\, \lt(\t\zeta_{\frc
  m{T_\rl}}(d+1)\,T_\rl^{d+1}+ \t\zeta_{\frc
  m{T_\rr}}(d+1)\,T_\rr^{d+1}\rt),
	\label{ssval}
\eeqa
where we define the functions
\beq\label{zetatext}
	\zeta_a(z) := \frc1{\Gamma(z)} \int_0^\infty \mathrm{d}p\,\frc{p^{z-1}}{
	e^{\sqrt{p^2+a^2}}-1},\quad
	\t\zeta_a(z) := \frc1{\Gamma(z)} \int_0^\infty \mathrm{d}p\,\frc{p^{z-2}\sqrt{p^2+a^2}}{
	e^{\sqrt{p^2+a^2}}-1}.
\eeq

\begin{figure}
\begin{center}
\includegraphics[width=16cm]{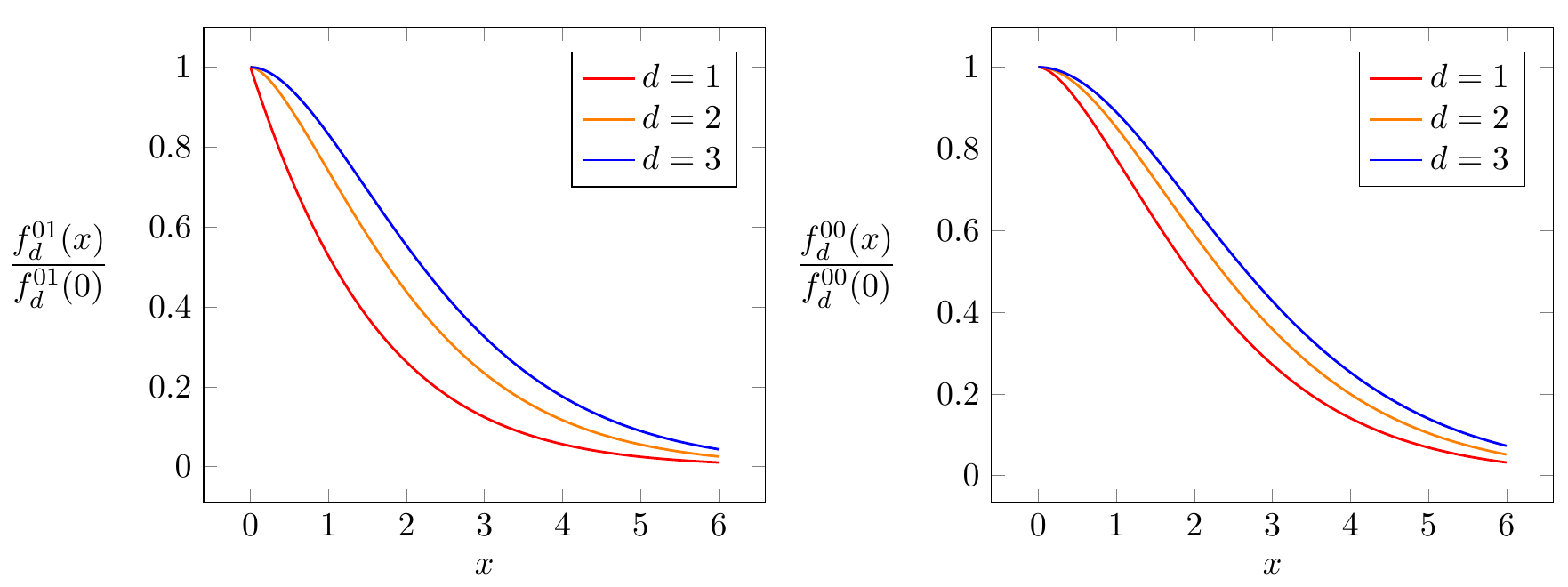}
\end{center}
\caption{Left: Energy current function, $f_d^{01}(x)$ ($x=m/T$) for $d=1,2,3$. Right:
  Energy density function, $f_d^{00}(x)$ for $d=1,2,3$. Both are exponentially decaying at large mass $m$ (low temperature $T$).}
\label{Fig:massfunctions}
\end{figure}

The averages in Eq.~\eqref{ssval} separate into sums and differences
of functions of $T_\rl$ and $T_\rr$, as is the case in 1+1 CFT
\cite{doyon1d1, doyon1d2} and free particle models
\cite{tas,ah,og,aschp,collura2}. This is a consequence of the
triviality of the scattering matrix and differs from the generic case
of higher-dimensional CFT \cite{bhaseen1}. In the massless limit
(corresponding to $a=0$) both functions in Eq.~\eqref{zetatext}
specialize to the Riemann zeta function, $\zeta_0(z) = \t\zeta_0(z) =
\zeta(z)$. In particular, setting $d=1$ and $m=0$, and using $\zeta(2)
= \pi^2/6$ with $\Gamma(1/2)=\sqrt{\pi}$, we find that the
coefficients of the powers of temperature all specialize to $\pi/12$
(both for $\bra T^{00}\ket_{\rm s}$ and $\bra T^{01}\ket_{\rm s}$) as
required for a 1+1 CFT with central charge $c=1$
\cite{doyon1d1,doyon1d2}. The $T^{d+1}$ temperature dependence
appearing in Eq.~\eqref{ssval} when $m=0$ is analagous to the
Stefan--Boltzmann law for black body radiation \cite{bhaseen1}. More
generally, in the massive case this temperature dependence is
multiplied by a function of $m/T$.  Setting $T_{\rm L}=T$ and $T_{\rm
  R}=0$ one may denote $\bra T^{01}\ket_{\rm s} =\bra T^{01}\ket_{\rm
  s}^{m=0}f_d^{01}(m/T)$ and $\bra T^{00}\ket_{\rm s} =\bra
T^{00}\ket_{\rm s}^{m=0}f_d^{00}(m/T)$. The functions $f_d^{01}(m/T)$
and $f_d^{00}(m/T)$ are plotted in Fig.~\ref{Fig:massfunctions} for
different spatial dimensions.

Results for the free massive Dirac model are presented in Appendix
\ref{appfermion}.

\subsection{Exact cumulant generating function of the energy transfer}

The observables of most interest in the general theory of
non-equilibrium steady states are the scaled cumulants of the quantity
being transferred. The generating function of these cumulants is
 related to the large-deviation function \cite{esposito}, which is
thought to be a good candidate for replacing the free energy as a
``thermodynamic potential'' in non-equilibrium systems. The cumulants
provide the full long-time statistics of the transferred quantity, and
hence much more information about the physics of the transfer process.

The fundamental definition of the scaled cumulants (here in the case
of energy transfer) is as follows. Suppose we measure the energy $q$
transferred from the left to the right a time $t$ after the
connection. In order to have finite energy transfer in finite time, we
assume that the $d-1$ dimensional transverse directions (transverse to
the flow) have linear size $L$. The quantity $q$ is a random quantity,
and we may represent the associated probability measure by
$\Omega_{t,L}$. According to standard statistical definitions, the
cumulants $\bra q^n\ket_{\Omega_{t,L}}^{\rm cumul}$ of the random
variable $q$ may be defined from its averages by the generating
function $\sum_{n=1}^\infty \frc{z^n}{n!} \bra
q^n\ket_{\Omega_{t,L}}^{\rm cumul} = \log \bra
\mathrm{e}^{zq}\ket_{\Omega_{t,L}}$. The scaled cumulants, and scaled
cumulant generating function (SCGF), are then defined as,
respectively, \beq c_n:=\lim_{t\to\infty\atop L\to\infty}
\frc1{tL^{d-1}} \bra q^n\ket_{\Omega_{t,L}}^{\rm cumul},\quad F(z) =
\sum_{n=1}^\infty \frc{z^n}{n!} c_n.  \eeq This assumes that the
cumulants of the variable $q$ scale proportionally to the time and to
the transverse area at large time and large transverse area.

In order to have a working definition of $c_n$, we would need to
define more precisely $\Omega_{t,L}$. The exact definition of the
measure $\Omega_{t,L}$ in quantum systems is subtle because, contrary
to the case of classical systems, it requires a precise description of
the quantum measuring process and of its influence on the system. Here
we will not discuss these subtleties - for discussions see for instance \cite{berjmp,doyon1d2}.

Instead, we will adopt a simple and intuitive expression for $c_n$,
which is frequently used in the literature and is expected to arise as a
result of various measurement protocols. This expression gives the
scaled cumulants in terms of the {\em connected correlation functions}
of the integrated energy current density. That is, cumulants are
expressed as connected correlation functions of the current density
$T^{01}(x,t) = T^{01}(x^1,\t x,t)$ integrated over the coordinates $\t x$
parametrizing the transverse direction and over time $t$, evaluated on
the connection hypersurface $x^1=0$: \beq\label{cn} c_n =
\lim_{\ep\to0^+} \int \prod_{j=1}^{n-1} \mathrm{d}^{d-1} \t x_{j}
\int_{-\infty}^\infty \prod_{j=1}^{n-1} \mathrm{d} t_{j}\, \bra T^{01}(0,\t
x_{n-1},t_{n-1}^{(\ep)})\cdots T^{01}(0,\t x_1,t_{1}^{(\ep)})
T^{01}(0,0,0)\ket_{\rm s}^{\rm conn.}  \eeq where $t^{(\ep)}_j = t_j
+ij\ep$. This expression should be interpreted as the scaled average
of the $n^{\rm th}$ power of the total 
energy
passing through the
hypersurface in a long time and for a large transverse area. One
regularizes the UV singularities arising from colliding fields by
imposing an imaginary time ordering to the factors: hence the
$\ep$-regularization in \eqref{cn}. The result is finite and real.

The integrals in Eq.~\eqref{cn} are in general hard to evaluate. One
can however evaluate the full SCGF directly by using the extended
fluctuation relations (EFR), derived in \cite{doyontrans} and shown
there to hold in free models of any dimensionality. The EFR state that
\beq\label{Fz} F(z) = \int_0^{z} \mathrm{d}y\,J_{{\rm E}}(\beta_\rl -
y,\beta_\rr + y) \eeq where $J_{{\rm E}}(\beta_\rl,\beta_\rr) =
\bra T^{01}\ket_{\rm s}$ is the steady-state energy current density as
a function of the inverse temperatures. In particular, the cumulants
can be evaluated in terms of derivatives of the current, thus
providing expressions for the non-trivial multiple integrals
\eqref{cn} directly from the exact expression \eqref{ssval}. Using the
mode expansion \eqref{modeexp} in the stress-energy tensor
\eqref{tmunu}, along with Wick's theorem and the contractions
\eqref{traa}, we have evaluated explicitly $c_2$ using \eqref{cn}, and
verified that it is in agreement with the coefficient of $z^2/2$ on
the right-hand side of \eqref{Fz}.

A convenient way of representing the SCGF that has a clear physical
meaning as the transport of energy quanta is via a sum of independent
Poisson processes. The SCGF for a single Poisson process representing
energy transfers by quanta $q\in\R$ is $\mathrm{e}^{zq}-1$, and the SCGF is in general additive for
independent processes.  Hence, we wish to express $F(z)$ as
\beq\label{Fomega} F(z) = \int \mathrm{d}q\,\omega(q)\,(\mathrm{e}^{zq}-1).  \eeq In
order for this to have an interpretation as a sum of independent
Poisson processes, the weight $\omega(q)$ must be positive. If this is
the case, then \eqref{Fomega} implies that the long-time scaled energy
transfer can be fully reproduced by a classical process whereby quanta
of energy between $|q|$ and $|q|+dq$, traveling towards the right
($q>0$) or the left ($q<0$) in a cross section $\mathrm{d}^{d-1}\t x$,
are distributed uniformly so that they cross the measuring
hypersurface with a flat probability weighted by $\mathrm{d}^{d-1}\t x\,\mathrm{d}q
\,\omega(q)$.

Following \cite{doyonint} we can express the weight as a Fourier
transform of the current, \beq \omega(q) = \frc1q \int
\frc{\mathrm{d}\lambda}{2\pi} \,J_{\rm E}(\beta_\rl -i\lambda,\beta_\rr
+i\lambda)\,\mathrm{e}^{-\mathrm{i}\lambda q}.  \eeq Using \eqref{T01} in the form \beq
J_{\rm E}(\beta_\rl,\beta_\rr) = \int \frc{\mathrm{d}^d p}{(2\pi)^d} |p^1|
\sum_{n=1}^\infty \lt( \Theta(p^1)\,\mathrm{e}^{-n\beta_\rl E_p} -
\Theta(-p^1)\,\mathrm{e}^{-n\beta_\rr E_p}\rt) \eeq and performing explicitly
the $\lambda$ and $p$ integrals, one obtains \beqa \omega(q) &=&
\sum_{n=1}^\infty \int \frc{\mathrm{d}p^2\cdots \mathrm{d}p^d}{(2\pi)^d\,n^2}
\lt(\mathrm{e}^{-\beta_\rl q}\, \Theta \lt(q-n\sqrt{\t p^2+m^2}\rt) +
\mathrm{e}^{\beta_\rr q}\, \Theta \lt(-q-n\sqrt{\t p^2+m^2}\rt) \rt) \n &=&
\lt(2^{d}\pi^{\frc{d+1}2} \Gamma\lt(\frc{d+1}2\rt)\rt)^{-1}\,
\sum_{n=1}^{[|q|/m]} \frc1{n^{d+1}}\lt(q^2-n^2m^2\rt)^{\frc{d-1}2}
\cdot\lt\{\ba{ll} \mathrm{e}^{-\beta_\rl q} & (q>0) \z \mathrm{e}^{\beta_\rr q} & (q<0)
\ea\rt. \label{omega} \eeqa where $\t p^2 = \sum_{i=2}^d (p^i)^2$ and
$[|q|/m]$ is the integer part of $|q|/m$. This is clearly positive,
hence the long-time fluctuations are correctly represented by a family
of independent Poisson processes. For $d=1$ the expression reduces to
that obtained for the one-dimensional Ising model \cite{doyonising}
except for the absence of the fermionic sign factor $(-1)^{n-1}$.

The SCGF can be evaluated using \eqref{Fomega} as an infinite
sum of modified Bessel functions: \beq F(z) = f(z,\beta_\rl) +
f(-z,\beta_\rr), \eeq where \beq\label{f} f(z,\beta) =
\lt(\frc{m}{2}\rt)^{\frc d2} \sum_{n=1}^\infty \frc{1}{(\pi n)^{1+\frc
    d2}} \lt( (\beta-z)^{-\frc d2} K_{d/2}(nm(\beta-z)) - \beta^{-\frc
  d2} K_{d/2}(nm\beta)\rt).\eeq
Note that the symmetry $\beta_\rl\mapsto \beta_\rr+z$, $\beta_\rr\mapsto \beta_\rl-z$ is a consequence of the EFR \eqref{Fz}.

In the massless limit \eqref{omega} yields \beq \omega(q)
\stackrel{m=0}= \frc{\zeta(d+1)}{2^{d}\pi^{\frc{d+1}2}
  \Gamma\lt(\frc{d+1}2\rt)} q^{d-1} \cdot\lt\{\ba{ll} \mathrm{e}^{-\beta_\rl q}
& (q>0) \z \mathrm{e}^{\beta_\rr q} & (q<0). \ea\rt. \label{omegam0} \eeq For
$d=1$ this reproduces the 1+1 CFT result \cite{doyon1d1, doyon1d2} for central charge
$c=1$. We see that the energy dependence of the weights for the
independent energy quanta are determined not only by the Boltzman
distribution, as in the $d=1$ case, but also by a factor $q^{d-1}$ due
to the impact of the extended transverse area. Using \eqref{omegam0}
the massless limit of the SCGF can be directly evaluated: \beq F(z)
\stackrel{m=0} = \frc{\Gamma\lt(\frc{d}2\rt)\zeta(d+1)}{2\pi^{1+\frc
    d2}} \lt((\beta_\rl-z)^{-d} + (\beta_\rr+z)^{-d} - \beta_\rl^{-d}
- \beta_\rr^{-d}\rt). \eeq
Explicitly, the cumulants in the massless limit are
\beq
	c_n = \frc{\Gamma(d+n) \zeta(d+1)}{2^d \pi^{\frc{d+1}2}\Gamma\lt(\frc{d+1}2\rt)}\lt(\beta_{\rl}^{-d-n} + (-1)^n \beta_{\rr}^{-d-n}\rt).
\eeq
Again, for $d=1$ this agrees with the known
1+1 CFT result \cite{doyon1d1, doyon1d2}. Results for the SCGF in the free massive Dirac models are presented in Appendix \ref{appfermion}.

\subsection{Non-locality of the steady-state density matrix and power-law correlations}

\renewcommand{\h}{\widehat}
The steady-state density matrix is described in Eq. \eqref{rhoness} in
terms of creation and annihilation operators. It has the form
$\rho_{\rm s} = \exp [- \h W]$, for a specified operator $\h
W$. Naturally, $\h W$ is conserved by the full dynamics, $[H,\h W]=0$,
as the density matrix represents a state that is stationary. It is
natural to ask whether $\h W$ may be written as one of the
infinitely-many local conserved charges of the Klein-Gordon
theory. The locality of $\h W$, or lack thereof, is important, as it
has implications for the large-distance decay of correlation functions
in the steady state.

We may evaluate $\h W$ explicitly in terms of the local fields
$\phi(x)$ and $\pi(x)$ by using \eqref{rhoness} and the inversion of
\eqref{modeexp}. The calculation is shown in Appendix
\ref{apploc}. The result may be expressed in the following form:
\beq\label{hW} \h W = \frc{\beta_\rl + \beta_\rr}2 \,H +
\frc{\beta_\rl-\beta_\rr}2 \,(P_1 + \h Q), \eeq where $P_1 = \int
d^dx\,\phi(x) \p_1\pi(x)$ is the momentum operator in the longitudinal
direction and \beq \h Q = \int \mathrm{d}^dx \mathrm{d}^dy\,:\phi(x) \pi(y):
Q(x-y). \eeq The kernel $Q(x-y)$ is given by \beq\label{kern} Q(x) = -\frc{{\rm
    sign}(x^1)}{\pi}\int \frc{\mathrm{d}^{d-1}\t p}{(2\pi)^{d-1}} \, \mathrm{e}^{\mathrm{i}\t
  p\cdot \t x} \int_0^{E_{\t p}} \mathrm{d}\ell\, {\cal E}_{\ell,\t p}
\,\mathrm{e}^{-|x^1|\,\ell}, \eeq where ${\cal E}_{\ell,\t p} = \sqrt{|\t p|^2
  +m^2 - \ell^2}$, $E_{\t p} = {\cal E}_{0,\t p}$, and the tilde-variables
are transverse coordinates.

In \eqref{hW} we recognize the terms $\frc{\beta_\rl + \beta_\rr}2 \,H
+ \frc{\beta_\rl-\beta_\rr}2 \,P_1$ as representing a Lorentz boost of
the Hamiltonian in the longitudinal direction. These terms alone would
give rise to a density matrix of a boosted thermal state with boost
velocity $\frc{T_\rl-T_\rr}{T_\rl+T_\rr}$ and rest-frame temperature
$\sqrt{T_\rl T_\rr}$. Remarkably, this has the same structure as the
exact non-equilibrium steady-state density matrix of a {\em
  one-dimensional CFT} \cite{doyon1d1,bhaseen1}, although $H$ and
$P_1$ pertain to the higher-dimensional massive system. 
The third term, involving $\h Q$, is a
correction to this, which further 
accounts for the higher dimensionality and
the non-zero mass. One can check that it indeed vanishes if and only
if $d=1$ and $m=0$. The third term is not an integration over a local
density, and in fact, the kernel connects local fields in the bilinear
expression over long distances as it does not decay exponentially. For
instance, in the case with $d=1$ and $m\neq 0$, the kernel has the
following large-$|x^1|$ asymptotic expansion: \beq Q(x) = -\frc{m}{\pi
  x^1}\,(1+O(1)).  \eeq For $d>1$, the decay of $Q(x)$ is also
$O(1/x^1)$ at large $|x^1|$, with a coefficient that involves both the
square-root of the transverse Laplacian $\sqrt{-\nabla_{\t
    x}^2}\;\delta^{(d-1)}(\t x)$, and a regular function of the
transverse coordinates $\t x$.

The algebraic decay of $Q(x)$ at large $|x^1|$ is the signature of
large-distance algebraic correlations. These are indeed known to exist
in non-equilibrium steady states \cite{zha,dor}, and are usually
attributed to the lack of detailed balance.
The algebraic decay can be seen explicitly in the following
correlation function in the one-dimensional case: \beq\label{phipis}
\bra\phi(x)\pi(0)\ket_{\rm s} \sim \frc1{4\pi |x^1|}
\frc{\sinh\frc{(\beta_\rl-\beta_\rr)m}2}{\sinh\frc{\beta_\rl m}2
  \sinh\frc{\beta_\rr m}2} \quad \mbox{as $|x^1|\to\infty$ \quad
  ($d=1$)}.  \eeq Interestingly, however, the correlation function
$\bra\phi(x)\phi(0)\ket_{\rm s}$ decays {\em exponentially} at large
distances, emphasizing the fact that the presence of the algebraic
decay depends on the observables involved. For further details of
these calculations see Appendix \ref{appcorr}.

Some comments are in order. First, we note that, at least in
one-dimensional systems, it has been known for some time that the
density matrix for non-equilibrium quantum steady states may be the
exponential of a non-local operator. For instance, in the context of
the non-equilibrium Kondo impurity model, Hershfield's density matrix
\cite{hers} was analyzed in \cite{doyand} and argued there to have a
non-local form. Non-locality was shown explicitly in \cite{doylect}
for the resonant-level impurity model. Recently, certain quantum
steady states in integrable spin chains were also shown to give rise
to density matrices of a non-local form \cite{mpp}.

Second, it is important to realize that non-equilibrium quantum steady states are not always described by density matrices of a non-local form, and do not always present algebraic correlations. Indeed, as a consequence of the results of \cite{doyon1d1} in one-dimensional CFT, and of \cite{bhaseen1} in interacting higher-dimensional CFT, the density matrix is the exponential of a local conserved charge, and correlations are exponentially decreasing in these cases. An exponential decay (at least close to equilibrium) was also proven mathematically for spin-spin correlation functions in the XY chain \cite{asch}, and derived physically in the one-dimensional Ising field theory \cite{chen}, despite
the fact that the density matrix has a 
non-local form in terms of the underlying fermions.

Finally, we remark that an interesting phenomenon occurs in
one-dimension: by the above discussion, we see that some correlations
decay algebraically in the steady state of the massive $d=1$
Klein-Gordon model, but exponentially in the massless model.

\section{Exact time evolution and the steady state density matrix}
\label{Sect:Time}

We now wish to evaluate explicitly the limit \eqref{limit} for local operators $\Or$, and show the form \eqref{ness} of the steady state with the density matrix \eqref{rhoness}. We will use techniques based on equations of motion developed in the context of free fermionic quantum impurity problems in \cite{berjmp}.

\subsection{$A$- and $B$-representations}

In the previous section, we have written the representation
\eqref{modeexp}, \eqref{crm} (which we will refer to as the
$A$-representation) of the canonical commutation relations
\eqref{cr}. This representation diagonalizes the Hamiltonian $H$ on
the line $x^1\in[-\infty,\infty]$, hence it is efficient in
order to evaluate time-evolved fields
$\phi(x,t)=\mathrm{e}^{\mathrm{i}Ht}\phi(x)\mathrm{e}^{-\mathrm{i}Ht}$ and
$\pi(x,t)=\mathrm{e}^{\mathrm{i}Ht}\pi(x)\mathrm{e}^{-\mathrm{i}Ht}$: \beqa \phi(x,t)&=& \int D p \,(A_p
\mathrm{e}^{-\mathrm{i}E_pt+\mathrm{i}p\cdot x} + A^\dag_p \mathrm{e}^{\mathrm{i}E_pt-\mathrm{i}p\cdot x}) \n \pi(x,t)&=&
-\mathrm{i}\,\int D p \,E_p\, (A_p \mathrm{e}^{-\mathrm{i}E_pt+\mathrm{i}p\cdot x} - A^\dag_p
\mathrm{e}^{\mathrm{i}E_pt-\mathrm{i}p\cdot x}).
	\label{modeexpt}
\eeqa
However, in this representation, the Hamiltonians $H_{\rm L}$ and $H_{\rm R}$ 
take a complicated (although still bilinear) form. Hence, evaluating the limit \eqref{limit} is a difficult task.

There is another representation (the $B$-representation) of the
commutation relations \eqref{cr}, which diagonalizes both $H_{\rm L}$
and $H_{\rm R}$. This is the representation used for describing two
boundary Klein--Gordon models, one on the negative half-line and the
other on the positive half-line, respectively. In the following, we
will take free boundary conditions at $x^1=0$, with the
condition $(\partial_{1}\phi)(0)=0$, and mention how the results are
modified for other boundary conditions. The mode expansion for the
Klein--Gordon model with free boundary conditions at $x^1=0$ is given
by \beqa \phi(x) &=& \int Dp\,(B_p \mathrm{e}^{\mathrm{i}\t p \cdot \t x} + B_p^\dag
\mathrm{e}^{-\mathrm{i}\t p \cdot \t x} )\,2\cos (p^1x^1) \Theta(-p^1x^1), \n \pi(x) &=&
-\mathrm{i} \int Dp\,E_p\,(B_p \mathrm{e}^{\mathrm{i}\t p \cdot \t x} - B_p^\dag \mathrm{e}^{-\mathrm{i}\t p \cdot
  \t x})\, 2\cos (p^1x^1) \Theta(-p^1x^1). \label{newmodeexp} \eeqa
Here, $\t p\cdot \t x = \sum_{i=2}^d p^ix^i$ is the dot product in the
transverse direction. In this representation, the operators $B_p$ and
$B_p^\dag$ satisfy the same canonical commutation relations as do
$A_p$ and $A_p^\dag$ \eqref{crm}, and we have \beq\label{Hlr} H_{\rm
  L,R} = \int_{p^1\gtrless 0} Dp\, E_p\,B_p^\dag B_p.  \eeq This
representation is efficient in order to evaluate averages under
$\rho_0$. We have in particular \beq \Tr \lt(\frak{n}[\rho_{0}] B_p
B^\dag_q\rt) = \frc{(2\pi)^d\,2E_p\,\delta^d(p-q)}{1-\mathrm{e}^{-W(p)}},\quad
\Tr \lt(\frak{n}[\rho_{0}] B_p^\dag B_q\rt) =
\frc{(2\pi)^d\,2E_p\,\delta^d(p-q)}{\mathrm{e}^{W(p)}-1}.
	\label{trbb}
\eeq Note the similarity with \eqref{traa}. However, in this
representation, it is much more complicated to evaluate the
time-evolved operators $\mathrm{e}^{\mathrm{i}Ht}\phi(x)\mathrm{e}^{-\mathrm{i}Ht}$ and
$\mathrm{e}^{\mathrm{i}Ht}\pi(x)\mathrm{e}^{-\mathrm{i}Ht}$.

In order to show convergence to the steady state, we will establish that
\beq\label{relS}
	\lim_{t\to\infty} \Tr\lt( \frak{n}[\rho_0] \,\mathrm{e}^{\mathrm{i}Ht}\Or\mathrm{e}^{-\mathrm{i}Ht}\rt)
	= \Tr\lt( \frak{n}[\rho_0] \,S(\Or)\rt),
\eeq
for any product of operators $\Or = \prod_i \Or_i$, where $\Or_i$ are formed by normal-ordered products of $\phi(x_i)$, $\pi(x_i)$ and their derivatives. Here $S$ is the scattering isomorphism defined by
\beq\label{Smat}
	S(A_p) = B_p,\quad S(A_p^\dag) = B_p^\dag,\quad
	S\lt(\prod_{p} A_p^{\eta_p}\rt) = \prod_p S(A_p^{\eta_p}).
\eeq
Along with \eqref{trbb}, \eqref{traa} and Wick's theorem, this indeed shows \eqref{ness} with \eqref{rhoness}.

\subsection{Time-evolved operators: main results}

Following the techniques used in \cite{berjmp}, which may be used for any
free model, we evaluate the time evolution with $H$ by explicitly
solving the equations of motion \beq\label{eom} \dot\phi = \pi,\quad
\dot\pi = (\nabla^2-m^2)\phi \eeq in terms of the initial
conditions on $\phi(x)$ and $\pi(x)$, and then we replace the
initial conditions by their $B$-representation
\eqref{newmodeexp}. Having time-evolved fields in the
$B$-representation, we may then readily evaluate \eqref{limit} using
\eqref{trbb}.  The resulting integration is rather technical and we
relegate it to Appendix \ref{appendixintegrals}.  We find \beq \phi(x,t)
= \lt\{\ba{ll} \mathrm{e}^{\mathrm{i}H_0t}\phi(x)\mathrm{e}^{-\mathrm{i}H_0t} & (|x^1|>t) \z S(\phi(x,t))
+ \Psi(x,t) & (|x^1|<t) \ea\rt.
	\label{result}
\eeq where \beq \Psi(x,t) := \lt(\int Dp \,\mathrm{e}^{\mathrm{i}\t p\cdot \t x} B_p\,
G(x^1,t;p)+ h.c\rt), \eeq $H_0$ is given by \eqref{H0} and $S$ is the map
defined by \eqref{Smat}. The map $S$ corresponds to the steady state, and as we will see, the correction $\Psi(x,t)$ describes the approach to the steady state.

The function $G(x^1,t;p)$ is the solution of a {\em
    one-dimensional} Klein-Gordon equation with mass-squared $|\t
p|^2+m^2$, which satisfies the {\em opposite} of the initial,
pre-quench boundary conditions, so that free is exchanged with
fixed.  An integral representation for initial free boundary
conditions is \beq G(x^1,t;p) = \frc{\mathrm{i}\,{\rm sign}(p^1)}\pi
\int_{-E_{\t p}}^{E_{\t p}} \mathrm{d}u \, \frc{ \sinh\lt({\cal E}_{u,\t
    p}\,x^1\rt)}{E_p+u} \mathrm{e}^{\mathrm{i}ut} \quad \mbox{(initially free at
  $x^1=0$)},
	\label{corrF}
\eeq
and for initial fixed boundary condition is
\beq
	G(x^1,t;p) = -\frc{|p^1|}\pi
	\int_{-E_{\t p}}^{E_{\t p}}
	\mathrm{d}u \, \frc{ \cosh\lt({\cal E}_{u,\t p}\,x^1\rt)}{(E_p+u){\cal E}_{u,\t p}}
	\mathrm{e}^{\mathrm{i}ut}\quad \mbox{(initially fixed at $x^1=0$)},
	\label{corrFf}
\eeq
where ${\cal E}_{u,\t p} = \sqrt{|\t p|^2 +m^2 - u^2}$ and $E_{\t p} = {\cal E}_{0,\t p}$.

Let us highlight the results with the following observations:
\begin{enumerate}
\item According to \eqref{result}, the time evolution with $H$ is exactly the same as that with $H_0$ whenever $|x^1|>t$. This is causality: beyond the light-wedge emanating from the space-time region $t=0$, $x^1=0$, it is not possible to distinguish between the dynamics generated by $H$ and that generated by $H_0$, as the information of the quench at $t=0,\,x^1=0$ is out of reach.
\item The correction \eqref{corrF} is {\em exactly zero} at
  $x^1=0$. This holds as an operator statement and subsists under the
  normal-ordering operation. This implies that if $\Or$ is a local
  operator at $x^1=0$, ultra-local in the longitudinal direction (not involving $x^1$ derivatives), then \beq \Tr\lt(\frak{n}[\rho_0] \Or(t)\rt) =
  \bra \Or\ket_{\rm s} \eeq for {\em every} $t>0$, with initially free
  conditions at $x^1=0$. That is, the steady state is reached {\em
    instantaneously} for such operators at $x^1=0$. This surprising
  fact is strongly connected to the choice of free boundary conditions
  before the quench. The interpretation is that with free boundary
  conditions, the fields at $x^1=0$ may freely reach their steady
  state limit, and they do so immediately.
  
  With initially fixed boundary
  conditions, a similar phenomenon occurs but for local operators at $x^1=0$ involving only fundamental fields with {\em single $x^1$ derivatives}. Indeed for such operators the correction \eqref{corrFf} vanishes. Fixed boundary conditions on $\phi$ correspond to free boundary conditions on $\p_1 \phi$, so that a similar interpretation holds. These observations explain why the behaviour of $G(x^1,t;p)$ at $x^1=0$ is the {\em opposite} of the initial, pre-quench boundary condition.  We note that in general, operators involving only even-derivative fundamental fields reach the steady state instantaneously on the connection hypersurface for initially free boundary conditions, and operators involving only odd derivatives do so for initially fixed boundary conditions.
\item Let $m=0$ and consider the Klein--Gordon field {\em integrated
  over the perpendicular direction}, $\int \mathrm{d}x^2\cdots \mathrm{d}x^d
  \phi(x,t)$. In this case, $\t p=0$. Since for $m=0$ we have $E_{\t p=0}=0$, then {\em
    both corrections \eqref{corrF} and \eqref{corrFf} vanish}. That
  is, the massless Klein--Gordon field integrated over the
  perpendicular region instantaneously reaches its steady state
  form. Appropriately normalized, correlation functions of products of
  such integrated fields are finite, and immediately reach their
  steady-state value as soon as they all lie in the light-wedge.  In
  fact, this is a special case of the general statement according to
  which such correlation functions are exactly described by an
  effective {\em one-dimensional} theory; see Appendix
  \ref{app1d}. Here it is the $d=1$ free massless boson, for which the
  1+1 CFT results can be used: an instantaneous steady state
  everywhere within the light-cone \cite{doyon1d1, doyon1d2}.
\end{enumerate}
Additionally, we see that if both $m=0$ and $d=1$, then $E_{\t p=0}=0$ and the corrections \eqref{corrF} and \eqref{corrFf} vanish. In this case, we again recover the 1+1-dimensional CFT result, according to which the steady state is instantaneously reached in the light-cone, for any boundary conditions.

The result \eqref{result} allows us to evaluate any time-evolved
average in the local quench problem. For instance, the
expression for the average energy current is reported in
Appendix \ref{appt01} in Eqs.~\eqref{avercud} and \eqref{avercu}.

\subsection{Long-time limit and approach to the steady state}

Consider an operator $\Or=\prod_i \Or_i(t_i,x_i)$ which is the product of local observables at different space-time points. Each observable $\Or_i(t_i,x_i)$ is a normal-ordered product of the field $\phi(t_i,x_i)$ and/or their space-time derivatives. If the correction $\Psi(x,t)=\int Dp \,\mathrm{e}^{\mathrm{i}\t p\cdot \t x} B_p\, G(x^1,t;p)+ h.c$ vanishes in the limit $t\to\infty$, then we have, from \eqref{result},
\beqa
	\lim_{t\to\infty} \Tr\lt(\frak{n}[\rho_0] \,\mathrm{e}^{\mathrm{i}Ht} \Or \mathrm{e}^{-\mathrm{i}Ht} \rt)
	&=&\Tr\lt(\frak{n}[\rho_0] S(\mathrm{e}^{\mathrm{i}Ht} \Or \mathrm{e}^{-\mathrm{i}Ht})\rt) \n
	&=&\Tr\lt(\frak{n}[\rho_0] \mathrm{e}^{\mathrm{i}H_0t} S(\Or )\mathrm{e}^{-\mathrm{i}H_0t} \rt)\n
	&=&\Tr\lt(\frak{n}[\rho_0] S(\Or )\rt).\no
\eeqa
That is, we have recovered the relation \eqref{relS}, from which follows \eqref{ness} with \eqref{rhoness}.

We now show that the correction $\Psi(x,t)$ provides asymptotically
vanishing corrections at large $t$ in the average
$\Tr\lt(\frak{n}[\rho_0] \,\mathrm{e}^{\mathrm{i}Ht} \Or \mathrm{e}^{-\mathrm{i}Ht} \rt)$. We concentrate
first on the case where local observables $\Or_i(t_i,x_i)$ are
normal-orderings of powers of $\phi(t_i,x_i)$ without any
derivatives. In order to evaluate the leading asymptotic correction,
we again use \eqref{result} (second line) and \eqref{trbb} along with
Wick's theorem. Note that in \eqref{result}, the operator
$\phi(t_i+t,x_i)$ is a sum of two contributions: its steady-state form
$S(\phi(t_i+t,x_i))$, and the correction $\Psi(t_i+t,x_i)$. In
applying Wick's theorem, we have to sum over products of Wick
contractions $\big(\phi(t_i+t,x_i),\phi(t_j+t,x_j)\big)$ (with
possibly $i=j$). In order to evaluate the leading order result
at large $t$, we consider terms where all Wick contractions are
between steady-state forms $S(\phi(t_i+t,x_i))$, except for one
contraction involving a single correction
$\Psi(t_j+t,x_j)$. Integrating over momenta and over $u$, there are
oscillatory factors $\mathrm{e}^{\pm \mathrm{i} E_p\,t}$ coming from the steady-state
form, and $\mathrm{e}^{\mathrm{i}ut}$ coming from the $G$-correction. By the method of
stationary phase, the leading large-$t$ result is obtained by
integrating around the region where the oscillation frequency
vanishes. Since $0<E_{\t p} \leq E_p$ for every $p$, this happens only
around $p^1=0$ (where $E_{\t p} = E_p$) and $u = \pm E_{\t p}$.

Consider the contraction between a positive-energy term in
$S(\phi(t_i+t,x_i))$ and a negative-energy term in
$\Psi(t_j+t,x_j)$. This is of the form, after the change of variable
$u\mapsto -u$, \beq\label{aos} \int Dp\, f(p)\,\mathrm{e}^{\mathrm{i}E_p t_i - \mathrm{i}p\cdot
  x_i+\mathrm{i}\t p\cdot \t x_j} \int_{-E_{\t p}}^{E_{\t p}} \mathrm{d}u\, \frc{
  \sinh\lt({\cal E}_{u,\t p}\,x_j^1\rt)}{E_p-u} \mathrm{e}^{\mathrm{i}(E_p-u)t}
\mathrm{e}^{-\mathrm{i}ut_j} \eeq where $f(p)$ is regular and non-zero at $p^1\to0$. We
change variables to $s,v$ with $s=(p^1)^2/(2E_{\t p})$ and $v=E_{\t
  p}-u$ and write \beq \int \mathrm{d}^dp \int ^{E_{\t p}} \mathrm{d}u \propto \int
\mathrm{d}^{d-1} \t p \int_0 \frc{\mathrm{d}s}{\sqrt{s}} \int_0 \mathrm{d}v \eeq where we only
indicate the integration limit for the region of interest. We now use
$E_p = E_{\t p} + s +O(s^2)$, so that $E_p-u = s+v +O(s^2)$, and
${\cal E}_{u,\t p} \propto \sqrt{v} (1+ O(v))$. Omitting
proportionality factors the leading large-$t$ behaviour of \eqref{aos}
is \beq\label{xt} x_j^1 \int_0 \frc{\mathrm{d}s}{\sqrt{s}} \int_0 \mathrm{d}v\,\sqrt{ v}
\,\frc{\mathrm{e}^{\mathrm{i} (s+v)t}}{s+v} \propto x^1_j t^{-1} \quad \mbox{(initially
  free at $x^1=0$).}  \eeq Similarly, using \eqref{corrFf} we get
\beq\label{xtf} \int_0 \mathrm{d}s \int_0 \frc{\mathrm{d}v}{\sqrt{ v}} \,\frc{\mathrm{e}^{\mathrm{i}
    (s+v)t}}{s+v} \propto t^{-\frc12} \quad \mbox{(initially fixed at
  $x^1=0$).}  \eeq In both cases, the proportionality factor is a
function of the mass, the temperatures, and $t_i$, $t_j$, $x_i$ and
$\t x_j$.

The results \eqref{xt} and \eqref{xtf} give the contribution to the
leading large-$t$ approach to the steady state coming from one Wick
contraction, and one choice of the member of the contracted pair where
the $\Psi$ term is taken. The leading large-$t$ asymptotic is obtained
by summing over every pair of fields, and for every pair, summing over
the two contributions coming from taking the $\Psi$ term for either
member in the pair. The contribution coming from the contraction
$\big(S(\phi(t_i,x_i)),\Psi(t_j,x_j)\big)$ is of the form $J\,x_j^1
t^{-1}$ (resp. $J\,t^{-1/2}$) for free (resp. fixed) initial boundary
condition, where the factor $J$ depends on $x_j^{2,\ldots,d}$,
$x_i^{2,\ldots,d}$, $t_i-t_j$, and on the other coordinates $x_k$,
$k\neq i,j$. The factor $J$ does not depend on $x_i^1$, $x_j^1$ or
$t_i+t_j$. Knowing this structure, we may extend this to local
observables containing $x_i$ and $t_i$ derivatives. The leading power
of $t$ is unchanged, unless a derivative with respect to $x_i^1$ of
high enough order makes the coefficient of the leading behaviour
vanish.

In conclusion, we find the following. {\em In a long-time asymptotic
  analysis, the steady state \eqref{ness} with \eqref{rhoness} is
  reached as $t\to\infty$, and is approached with a power law in $t$
  as follows: \beq \Tr\lt( \frak{n}[\rho_0]
  \,\mathrm{e}^{\mathrm{i}Ht}\Or \mathrm{e}^{-\mathrm{i}Ht}\rt) =
  \bra\Or\ket_{\rm s} + \lt\{\ba{ll} O\lt(t^{-1}\rt) &
  \mbox{(initially free at $x^1=0$)} \z O\lt(t^{-\frc12}\rt) &
  \mbox{(initially fixed (or mixed) at $x^1=0$)} \ea \rt.  
\label{statement} \eeq for
  any $\Or=\prod_i \Or_i(t_i,x_i)$ that is a product of local
  observables. Further, the coefficient in the initially free case
  decreases like $O(R)$ as $R\to0$, where $R = {\rm
    max}(\{|x_i^1|:i\})$ is the largest distance of a local observable
  to the connection hypersurface: \beq \lim_{t\to\infty} t\lt(
  \Tr\Big( \frak{n}[\rho_0] \,\mathrm{e}^{\mathrm{i}Ht}\Or
  \mathrm{e}^{-\mathrm{i}Ht}\rt) - \bra\Or\ket_{\rm s} \Big)
  \stackrel{R\to0}= O\big(R\big) \quad \mbox{(initially free at
    $x^1=0$)}.  \eeq } From this we make the following final
observation.  \bi
\item[4.] For free initial boundary conditions at $x^1=0$, the approach
  to the steady state, proportional to $t^{-1}$, is generically faster than
  that for fixed or mixed initial boundary conditions, proportional to
 $t^{-1/2}$. Again this may be explained as in
  Point 2 above, from the intuition that with free conditions, it is
  easier for the fields to reach their steady state value. They do so
  instantaneously at $x^1=0$, and faster the closer they are to the
  hypersurface $x^1=0$.  \ei

It is important to emphasize that equation (\ref{statement})
  gives the {\em leading} time dependence for generic observables. However,
  depending on the observable being considered, the leading order
  contribution may vanish giving rise to a faster approach.  As we
  will show in the next section, this occurs for the energy current
  and the energy density. Nonetheless, the approach to the steady state
  remains a power-law in time and is fully described by the above
  formalism.
  
An example of an observable whose leading time dependence is of the generic form is the two-point function of the fundamental field $\phi(x)$. For instance, an analysis as above, keeping track of the precise coefficients, give for initially fixed boundary conditions,
\beq
	\bra \phi(x,t)\phi(y,t)\ket -
	\bra \phi(x,t)\phi(y,t)\ket_{\rm s}
	= \frc1{\sqrt{\pi mt}}\,\frc{\sinh \frc{(\beta_\rl + \beta_\rr )m}2}{
	\sinh\frc{\beta_\rl m}2 \sinh\frc{\beta_\rr m}2}\lt(1
	+ O(t^{-1})\rt).
\eeq

\subsection{Time-evolution of the average energy current and energy density}
\label{Sect:EV}
\begin{figure}
\begin{center}
\includegraphics[width=16cm]{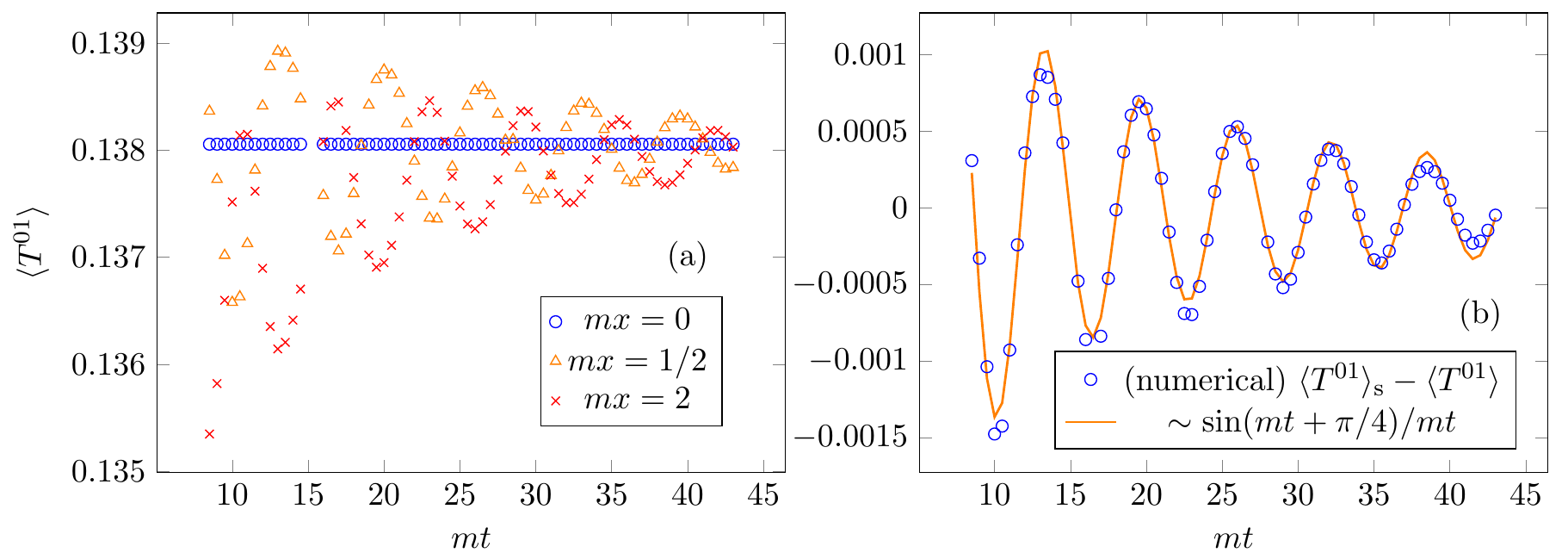}
\end{center}
\caption{Numerical solutions in $d=1$ with $T=m=1$ and free pre-quench
  boundary conditions. (a) Time-evolution of the energy current
  $\langle T^{01}\rangle$ at different distances $x$ from the
  connection hypersurface, showing the approach to the steady state
  $\langle T^{01}\rangle_s$ at late times, and the instantaneous
  approach at $x=0$.  (b) Case $x=1/2$. Fit to the function
    $\sin(mt+\pi/4)/mt$, giving a proportionality constant of $0.01395$.}
\label{Fig:T01free}
\end{figure}

\begin{figure}
\begin{center}
\includegraphics[width=16cm]{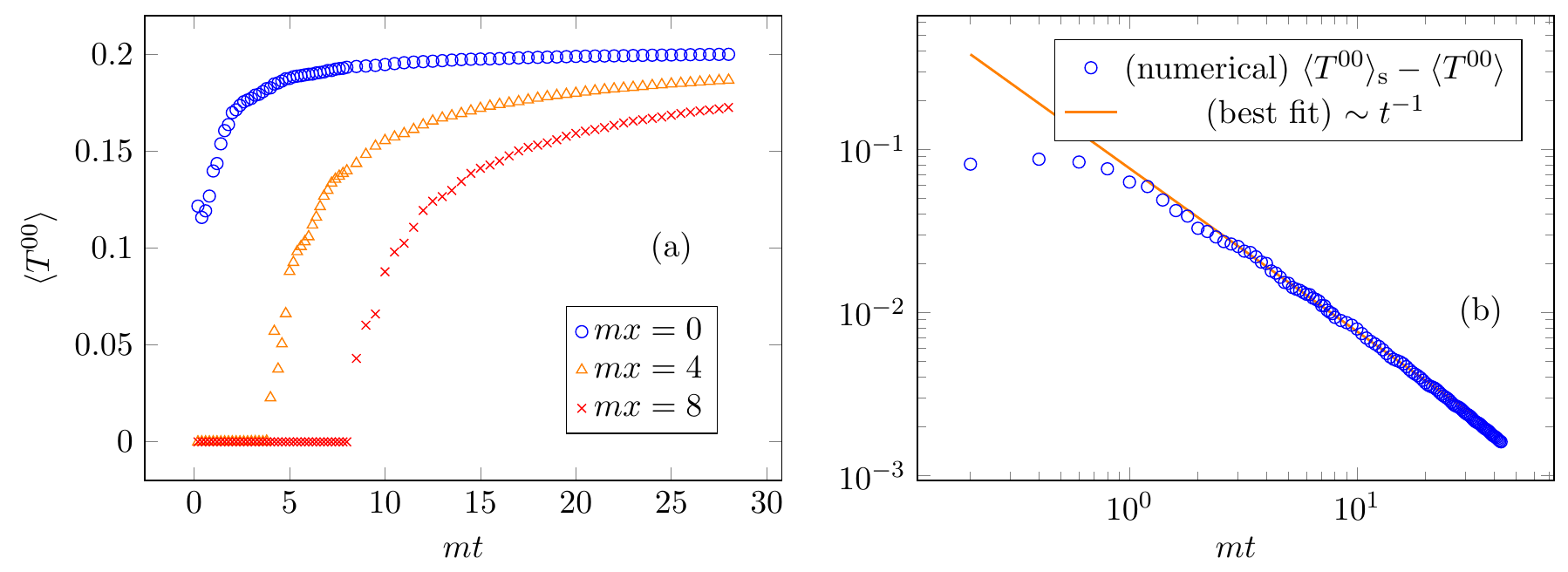}
\end{center}
\caption{Numerical solutions in $d=1$ with $T=m=1$ and fixed
    pre-quench boundary conditions. (a) Time-evolution of the energy
    density $\langle T^{00}\rangle$ at different distances $x$ from
    the connection hypersurface, showing the approach to the steady
    state $\langle T^{00}\rangle_s$ at late times. (b) The asymptotics are 
compatible with a $t^{-1}$ power-law approach. Here, the case $x=1/2$ is shown, where a proportionality constant of $0.07639$ is found.}
\label{Fig:T00fixed}
\end{figure}

\begin{figure}
\begin{center}
\includegraphics[width=16cm]{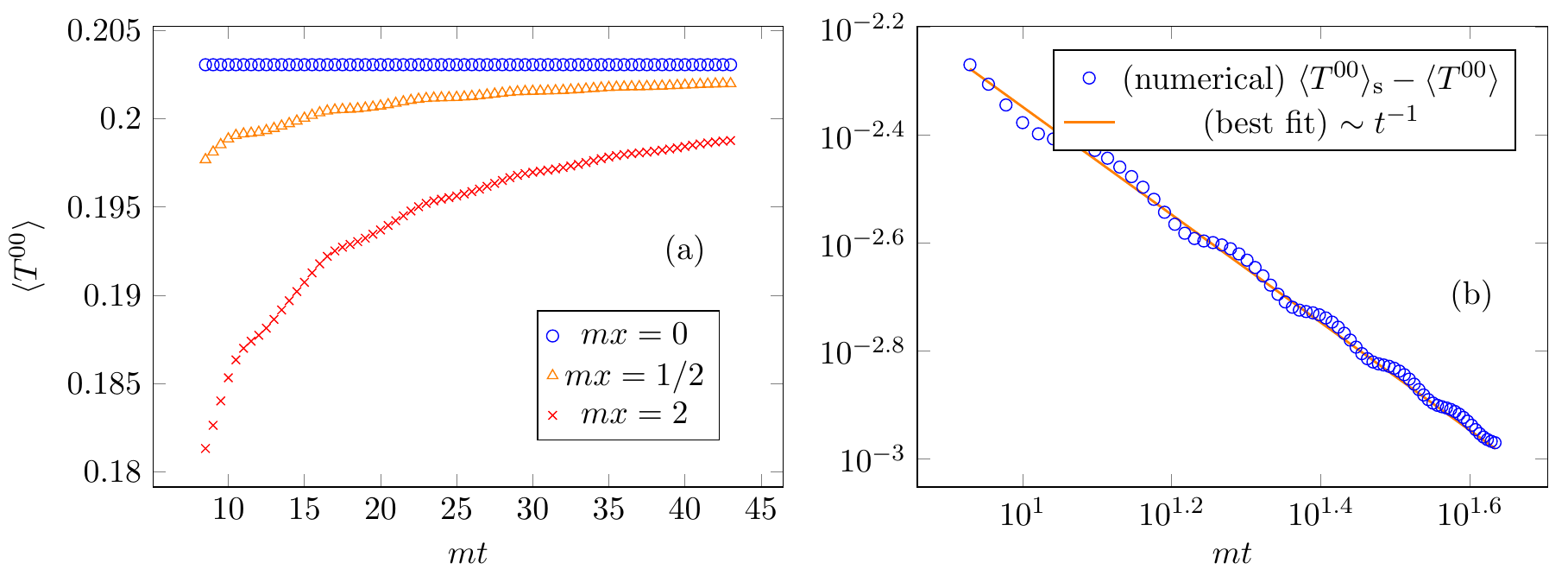}
\caption{Numerical solutions in $d=1$ with $T=m=1$ and free pre-quench
  boundary conditions. (a) Time-evolution of the energy density
  $\langle T^{00}\rangle$ at different distances $x$ from the
  connection hypersurface, showing the approach to the steady state
  $\langle T^{00}\rangle_s$ at late times, and the instantaneous
  approach at $x=0$.  (b) The asymptotics are compatible with a
$t^{-1}$ power-law approach (up to small oscillations). Here, the case $x=1/2$ is shown, where a proportionality constant of $0.0449$ is found.}
\label{Fig:T00free}
\end{center}
\end{figure}

\begin{figure}
\begin{center}
\includegraphics[width=16cm]{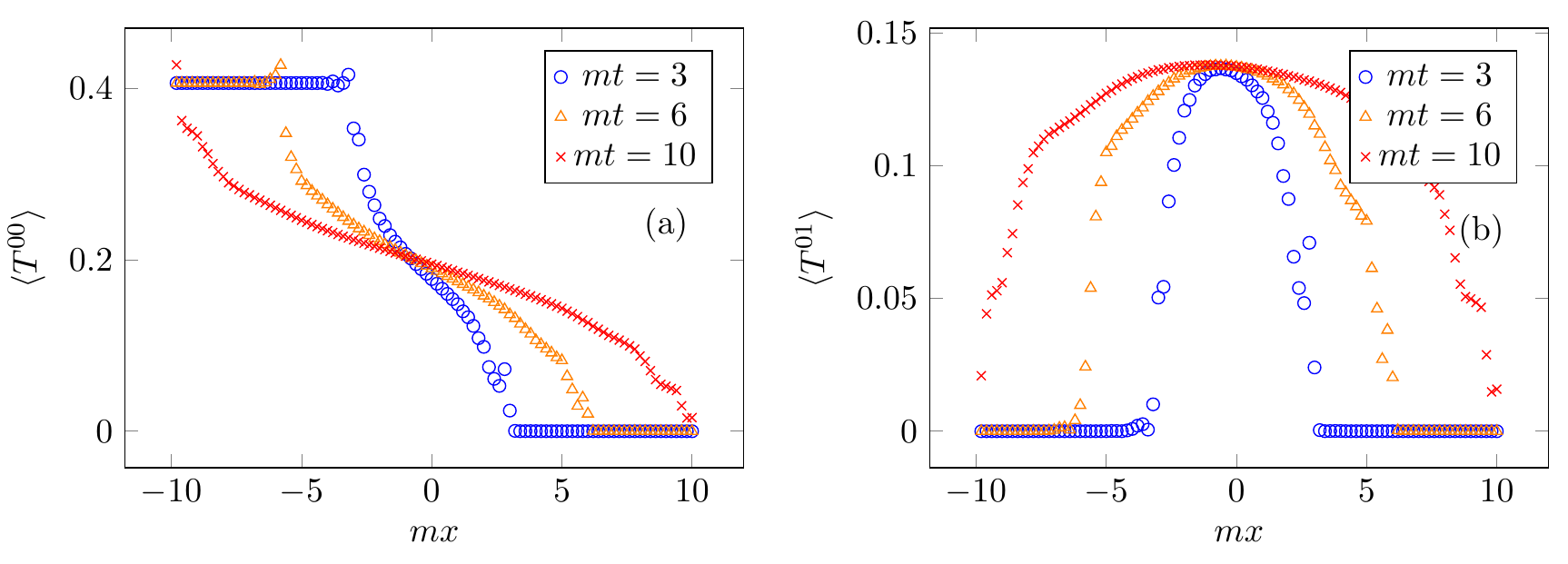}
\end{center}
\caption{Numerical solutions in $d=1$ with $T=m=1$ and fixed 
pre-quench boundary conditions. Spatial profiles of (a) the 
local energy density $\langle T^{00}(x)\rangle$ and (b) the energy current 
$\langle T^{01}(x)\rangle$ for different times.}
\label{Fig:Spatial}
\end{figure}

In the above analysis we have discussed the time-evolution of local
operators.  In this section we discuss the implications for the
time-evolution of the average energy current. For simplicity we
consider the case with $d=1$ and $m\neq 0$. As we discuss more fully
in Appendix \ref{appt01}, the long time approach to the NESS for fixed
initial boundary conditions is given by
\begin{equation}
\bra T^{01}\ket= \bra T^{01}\ket_s-\delta(\beta_{\rm L},\beta_{R}),
\label{jeasym}
\end{equation}
where $\delta(\beta_{\rm L},\beta_{R})\equiv \delta(\beta_{\rm
  L})-\delta(\beta_{R})$ and
\begin{equation}\label{delta}
\delta(\beta)= \frc{3\beta m}{64\pi \sinh^2(\beta m/2)} \lt(1
	+ \frc23\frc{\left(1-\mathrm{e}^{-\beta m}\right)}{\beta m}\sin(2mt)\rt)t^{-2} 
	+O(t^{-3}).
\end{equation}
The approach to the steady state is therefore governed by a $t^{-2}$
power-law with an oscillatory contribution.  This dependence is
confirmed numerically in Fig.~\ref{Fig:T01fixed}; 
for details of the numerical approach see  Appendix \ref{App:Numerics}. 
The corresponding
evolution for free initial boundary conditions is shown in
Fig.~\ref{Fig:T01free} and is compatible with a $t^{-1}$ approach.
Plots of the time evolution of the local energy density $\langle
T^{00}\rangle$ are shown in Figs.~\ref{Fig:T00fixed} and
\ref{Fig:T00free}. Plots of the time evolution of the spatial profile are given in
Fig.~\ref{Fig:Spatial}.

From Figs.~\ref{Fig:T01free}--\ref{Fig:T00free}, we observe that, for initially free boundary conditions, both the energy current and energy density approach the steady state with the generic (i.e.~the slowest allowed) power law $\propto t^{-1}$. On the other hand, for initially fixed boundary conditions, the approach is, in both cases, faster than the generic behaviour $\propto t^{-1/2}$. In all cases, the approach is modulated by oscillations, which appear to be stronger for the energy current than for the energy density. It is also interesting to note that for the energy current, initially fixed boundary conditions lead to a {\em faster approach} to the steady state than initially free boundary conditions, in contrast with the generic case. This could be understood by the fact, emphasized above, that initially fixed (free) boundary conditions for the fundamental field $\phi$ at $x^1=0$ correspond to initially {\em free (fixed)} boundary conditions for the longitudinal derivative $\p_1\phi$ at $x^1=0$. This naturally affects the longitudinal current, which reaches its steady state value more easily for initial conditions that are free on the longitudinal derivative $\p_1\phi$, hence fixed on $\phi$.

From Fig. \ref{Fig:Spatial} we observe the monotonic spacial behaviour of the energy density, interpolating between the left and right reservoirs; and the monotonicity of the energy current in the transition regions, interpolating between the reservoirs and the central region (where the steady-state forms). We also see the light-wedge effect, by which the average values of observables are unchanged beyond the light wedge $|x^1|>t$, in agreement with the first line of \eqref{result}.

\begin{figure}
\begin{center}
\includegraphics[width=16cm]{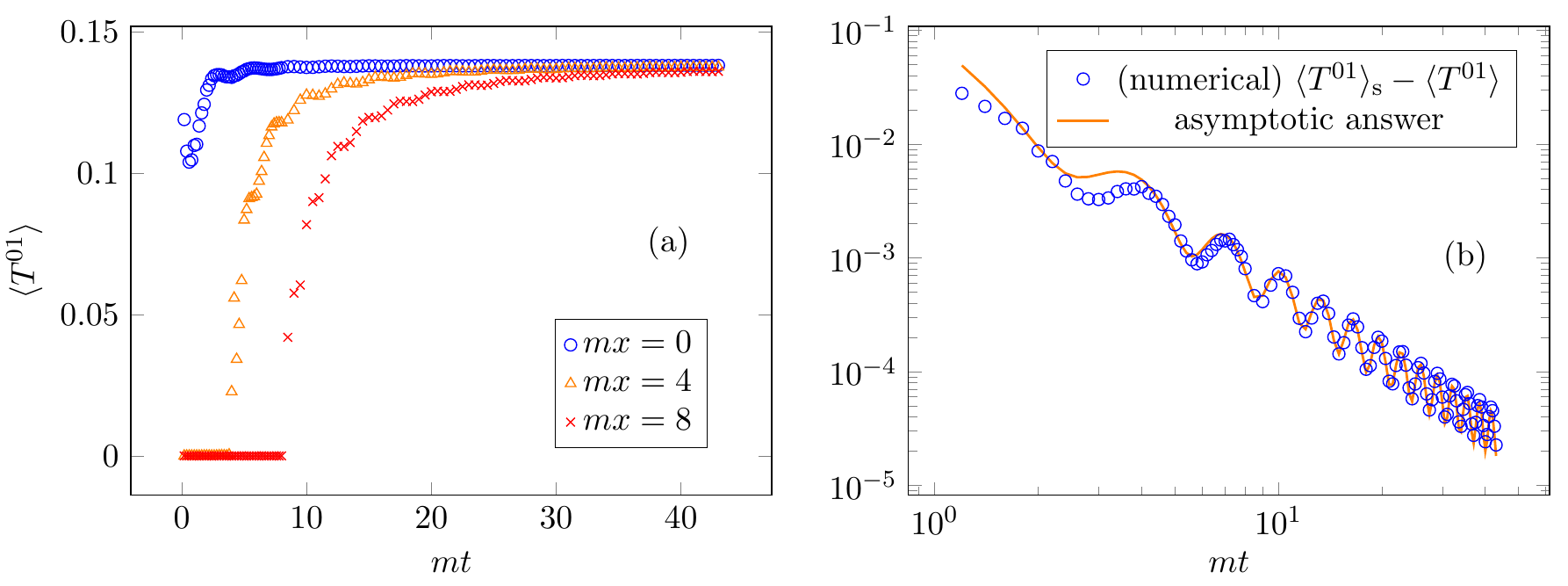}
\end{center}
\caption{Numerical solutions in $d=1$ with $T=m=1$ and fixed
  pre-quench boundary conditions. (a) Time-evolution of the energy
  current $\langle T^{01}\rangle$ at different distances $x$ from the
  connection hypersurface, showing the approach to the steady state
  $\langle T^{01}\rangle_s$ at late times. (b) Fit to the theoretical
  prediction (\ref{jeasym}) showing the oscillatory behavior and the
  $t^{-2}$ power-law.}
\label{Fig:T01fixed}
\end{figure}

\subsection{Semiclassical analysis of energy flow evolution and 
transition regions}

We have derived the properties of the steady state, including exact
expressions for the operator $\phi(x,t)$ and therefore for any
observable such as $T^{\mu\nu}(x,t)$.  It is also instructive to
understand how the steady state emerges in the simultaneous limit of
large time \emph{and} far from the connection region, by explicitly
evaluating expectation values of $T^{01}(x,t)$ using a semiclassical
analysis. This analysis is valid asymptotically in the limit where
$x^1,t\rightarrow \infty$ with $x^1/t$ fixed. Note that this is
outside of the regime of validity of Eq.~(\ref{corrF}). Here we
perform this calculation in the massless limit $m=0$ for simplicity,
but the procedure straightforwardly generalizes to the massive case.

The calculation is based on semiclassical insight into the evolution
of $\langle T^{\mu\nu}\rangle$.  A similar semiclassical approach was
used in \cite{collura1, collura2}. As this is a free theory, let us
begin by considering only the modes with momentum $p$ in the left
bath.  Without loss of generality, we may assume that $x^1>0$.  All of
the bosons with momentum $p$ move at the (group) velocity $p^1/|p|
\equiv v_p$ in the $x^1$-direction. On the world-surface $x^1/t=
v_{p_0}$, only modes at velocities $v_p>v_{p_0}$ will contribute from
the left as other modes are not fast enough to reach the
world-surface. On the other hand, from the right, all modes at
velocities $v_p<v_{p_0}$ will contribute. This includes all modes from
the right with negative velocity, and also some with positive
velocity, going in the ``wrong'' direction: those that are slow enough
to cross the world-surface. But modes from the right with opposite
velocities give exactly opposite contributions to the current, hence
cancel out. That is, the heat flow is given by counting all of the
momentum modes coming from opposite baths at velocities
$|v_p|<v_{p_0}$: positive velocity from the left, negative from the
right. An integral, mimicking the exact steady-state calculation described in 
Appendix \ref{appcal}, then gives us\begin{align} 
\label{semclass}
\langle
T^{01}\rangle &= (d-1)C_{\mathrm{L,R}}
\left(T_{\mathrm{L}}^{d+1}-T_{\mathrm{R}}^{d+1}\right)
\int\limits_0^{\theta_0} \mathrm{d}\sin \theta_1 \; \sin^{d-2}\theta_1
\notag \\ &= C_{\mathrm{L,R}}
\left(T_{\mathrm{L}}^{d+1}-T_{\mathrm{R}}^{d+1}\right)
\left(1-\frac{(x^1)^2}{t^2}\right)^{(d-1)/2},
\end{align}
where the normalization constants $C_{\mathrm{L,R}}$ are given in
Appendix \ref{appcal}, and $C_{\rm L} = C_{\rm R}$ in the massless case.  It is
readily seen that the same expression holds for $x^1/t<0$.
The above semiclassical argument may be justified from the exact
quantum computation; see Appendix \ref{appsemi}.

This semiclassical analysis emphasizes the fact that, as suggested by
Fig. \ref{Fig:Quench}, the transition region, interpolating between the baths and the
steady-state region, is large in $d>1$ and presents a smoothly
interpolating average current. We expect this to hold for other
observables as well, such as the energy density. The intuition is that modes of various velocities, which do not interact with each other, arrive, at any given point, at different times, slowly building up the steady state and producing large transients and transition regions. This is in strong
contrast with the $d=1$ results, where sharp transitions are found
between the asymptotic baths and the steady state, due to the one-dimensional relativistic dispersion relation giving rise to a single velocity (the velocity of light) for all momenta. 
The present
situation is very different: the transition region is large on
macroscopic 
scales.

The above semiclassical analysis is based on a picture of
freely-propagating particles. The evolution of $\langle T^{01}\rangle$
in a strongly interacting theory is inherently different. With
interactions, the time evolution of averages is quickly governed by
hydrodynamics, as interactions give rise to local thermalization
effects. Using these insights, it was found in \cite{bhaseen1} that at
leading order as $x^1/t\rightarrow \infty$, $\langle T^{01}\rangle$ is
a piecewise constant function with two jumps, corresponding to the
propagation of ``shock waves".

\section{Discussion: effects of interactions}

As we stressed in the introduction, in contrast to the one-dimensional
case, the steady state in the massless Klein--Gordon theory in $d>1$
is different from that found in interacting higher-dimensional CFT
\cite{bhaseen1}. This difference is due to the infinite number of
local conserved charges in the Klein--Gordon theory.  These conserved
charges imply that the occupation numbers $N_p = A^\dagger_p A_p$ are
conserved
for every momentum $p$, which allows for the momentum modes to be
thermalized independently. In a generic interacting QFT, 
one does \emph{not} find this large number of conserved
charges. Generically, one does not expect any conserved quantities
other than those associated with space-time symmetries: the total
energy $H$, the total momentum $P^i$, and the boost
  generators. By time and translation invariance of the steady state density matrix, the boost generator cannot contribute, so that in this case the non-equilibrium steady state is given
by a boosted thermal state \cite{bhaseen1}, rather than a
  collection of independently thermalized propagating modes.
This indicates that in, contrast to the situation in one-dimension 
where the free-field and boost descriptions coincide, in
higher-dimensions, the free-field NESS is 
{\em unstable} to perturbations. It would be very interesting to
understand the detailed evolution between the non-interacting and
interacting regimes. One possibility in a weakly interacting system is
that the non-equilibrium behavior is described by the free-field limit
at early times, before crossing over to the hydrodynamic regime. In
this scenario, the system would `pre-thermalize' as an ensemble of
independent and approximately conserved modes, followed by a more
rapid onset towards hydrodynamics. In this picture, the
pre-thermalization is approached as a power law in time, and would not
be seen unless the interactions are sufficiently weak so that the
hydrodynamic onset occurs at late enough times. An estimate of the
timescale for ``hydronization'' may be obtained by
dimensional analysis and scaling arguments. For the case of
``$\phi^4$" interactions:
\begin{equation} H_{\mathrm{int}} = \frac{\lambda}{4!} \int
  \mathrm{d}^dx\; : \phi(x,t)^4:
\end{equation}
one obtains
\begin{equation} t_{\mathrm{hydro}} \sim T^{5-2d}\lambda^{-2}.
\end{equation}
This formula is only valid when the dimensionless combination $\lambda
T^{d-3} \ll 1$, corresponding to the limit where kinetic theory is
valid; this is at large temperatures if $d<3$ (the coupling is
relevant and the free UV fixed point is nearby). At the marginal
dimension $d=3$ one finds the strongly-coupled, hydrodynamic form
$t_{\mathrm{hydro}} \sim 1/T$, up to a factor of $\lambda^{-2}$. It
would interesting to develop kinetic theory approaches to investigate
this crossover.

\section{Conclusion}
\label{Sect:Conc}

We have considered the non-equilibrium dynamics of the Klein-Gordon
model following a local quench in arbitrary dimension. We have adopted
the partitioning approach in which two independently thermalized
halves are brought into contact and are allowed to evolve
unitarily. We have demonstrated that the steady state density matrix
generically contains non-local contributions, away from the massless
limit in one-dimension.  This results in a power-law approach to the
steady state, where the exponent depends on the pre-quench boundary
conditions on the connection hypersurface. We provide exact results
for the steady state energy current, the energy density and the scaled
cummulants of the energy transfer. The statistics of the energy
transfer are described by a continuum of independent Poisson
processes, where the weights are analytically determined.
When specialized to zero mass and one dimension, the results found
here reproduce the known results from conformal field theory
\cite{doyon1d1, doyon1d2} at central charge $c=1$. Some of the results
are also in correspondence with those obtained in free fermion models
in higher dimensions \cite{collura2}.  However, in contrast to the
behavior of massless free-fields in one-dimension, which have a direct
connection to the generic results of $1+1$ CFT, the free-field limit
does {\em not} capture the non-equilibrium behavior of generic CFTs in
higher dimensions \cite{bhaseen1}. This is a direct consequence of the
anomalous behavior of free-fields due to the proliferation of
conservation laws. Many of these conservation laws are explicitly
broken in the presence of interactions and cannot contribute to the
generic non-equilibrium density matrix. It is interesting to note that
in the special case where $d=1$ and $m=0$ these additional
contributions drop out from the free-field density matrix so that the
free-field and CFT descriptions coincide. It would be very illuminating to understand
  the evolution between the free-field limit and the interacting
  regime in more detail for higher dimensional situations.

\section*{Acknowledgements}
\addcontentsline{toc}{section}{Acknowledgements}
 B.D. thanks the informal CFT discussion group at King's for comments on this work, \'Edouard Boulat, Pasquale
Calabrese, Mario Collura, M\'arton Kormos, Gabriele Martelloni, Hubert
Saleur and Jacopo Viti for discussions, and Universit\'e Paris
Diderot, where part of this work was done, for financial support
through a visiting professorship.  A.L. is supported by the Smith
Family Graduate Science and Engineering Fellowship, and would like to
thank the Perimeter Institute of Theoretical Physics for hospitality
as this work was in progress.  Research at Perimeter Institute is
supported by the Government of Canada through Industry Canada and by
the Province of Ontario through the Ministry of Economic Development
\& Innovation.  This work was further supported in part by a VICI
grant of the Netherlands Organization for Scientific Research (NWO),
by the Netherlands Organization for Scientific Research/Ministry of
\mbox{Science} and Education (NWO/OCW) and by the Foundation for
Research into Fundamental Matter (FOM). MJB acknowledges helpful
  discussions with Joel Moore and thanks the Thomas Young Center.

\appendix

\section{Energy current and energy density} \label{appcal}

Here we compute the  momentum current in the  steady state
\beq\label{T01app} \bra T^{01}\ket_{\rm s} = \int \frc{\mathrm{d}^d
  p}{(2\pi)^d} |p^1| \lt( \frc{\Theta(p^1)}{\mathrm{e}^{\beta_\rl E_p}-1} -
\frc{\Theta(-p^1)}{\mathrm{e}^{\beta_\rr E_p}-1}\rt).  
\eeq
To evaluate the $p_1$ integral we separate it into a sum of integrals over the
positive and negative real line, we find
\[
	\bra T^{01}\ket = C_\rl T_\rl^{d+1}-C_\rr T_\rr^{d+1},\quad
	\bra T^{00}\ket = A_\rl T_\rl^{d+1}+A_\rr T_\rr^{d+1}
\]
where
\beqa
	C_{\rl,\rr} &=& \int_{p_1>0} \frc{\mathrm{d}^dp}{(2\pi)^d} \frc{p_1}{\mathrm{e}^{\sqrt{p^2+m_{\rl,\rr}^2}}-1}
	= d! \,\zeta_{m_{\rl,\rr}}(d+1) \int_{|a|=1\atop a_1>0} \frc{\mathrm{d}^{d-1} a}{(2\pi)^d}\,a_1 \n
	A_{\rl,\rr} &=& \int_{p_1>0} \frc{\mathrm{d}^dp}{(2\pi)^d} \frc{\sqrt{p^2+m_{\rl,\rr}^2}}{\mathrm{e}^{\sqrt{p^2+m_{\rl,\rr}^2}}-1}
	=  d! \,\t\zeta_{m_{\rl,\rr}}(d+1) \int_{|a|=1\atop a_1>0} \frc{\mathrm{d}^{d-1} a}{(2\pi)^d} \no
\eeqa
with $m_{\rl,\rr} := m/T_{\rl,\rr}$ and
\[
	\zeta_b(d) :=\frc1{\Gamma(d)} \int_0^\infty \mathrm{d}p\,\frc{p^{d-1}}{\mathrm{e}^{\sqrt{p^2+b^2}}-1},\quad
	\t\zeta_b(d) :=\frc1{\Gamma(d)} \int_0^\infty \mathrm{d}p\,\frc{p^{d-2}\sqrt{p^2+b^2}}{\mathrm{e}^{\sqrt{p^2+b^2}}-1}.
\]
We note that $\zeta_0(d) = \t\zeta_0(d) = \zeta(d)$ is Riemann's zeta function. In order to evaluate the angular integrals, one may use the $d-1$ angles $\theta_1,\ldots,\theta_{d-1}$ on the hypersphere, with $a_1 = \cos\theta_1$. The total surface of the $(d-1)$-dimensional unit sphere (embedded in $d$ space dimensions) is
\[
	\int_{-\pi}^{\pi} \mathrm{d}\theta_{d-1}
	\int_0^\pi \mathrm{d}\theta_{d-2}\cdots \mathrm{d}\theta_1\,
	\sin\theta_{d-2}\sin^2\theta_{d-3}\cdots
	\sin^{d-2}\theta_1
	= \frc{2\pi^{d/2}}{\Gamma(d/2)}.
\]
Hence we directly have
\[
	A_{\rl,\rr} = \frc{d \,\Gamma\lt(\frc{d+1}2\rt) \t\zeta_{m_{\rl,\rr}}(d+1)} {2\pi^{(d+1)/2}}
\]
and we may calculate
\beqa
	\int_{|a|=1\atop a_1>0} \frc{\mathrm{d}^{d-1}a}{(2\pi)^d}\,a_1 &=&
	\frc1{(2\pi)^d}
	\int_{-\pi}^{\pi} \mathrm{d}\theta_{d-1}
	\int_0^\pi \mathrm{d}\theta_{d-2}\cdots \int_0^{\pi/2}\mathrm{d}\theta_1\,
	\sin\theta_{d-2}\sin^2\theta_{d-3}\cdots
	\sin^{d-2}\theta_1 \cos\theta_1 \n
	&=& 	\frc1{(2\pi)^d}
	\frc{2\pi^{(d-1)/2}}{\Gamma((d-1)/2)}
	\int_0^{\pi/2}
	\mathrm{d}\theta_1 \,\cos\theta_1\,\sin^{d-2}\theta_1 \n
	&=& 	\frc1{(2\pi)^d}
	\frc{2\pi^{(d-1)/2}}{\Gamma((d-1)/2)}
	\int_0^1
	\mathrm{d}v\, v^{d-2} \n
	&=& 	\frc1{(2\pi)^d}
	\frc{2\pi^{(d-1)/2}}{\Gamma((d-1)/2)}
	\frc1{d-1} \no
\eeqa
which gives
\beqa
	C_{\rl,\rr} &=& 
	\frc{d\,\Gamma\lt(\frc d2\rt)\zeta_{m_{\rl,\rr}}(d+1)}{2\pi^{d/2+1}}.
\eeqa
In $d=1$, using $\zeta(2) = \pi^2/6$ and $\Gamma(1/2)=\sqrt{\pi}$, we find $A=C=\pi/12$ as it should for a one-dimensional CFT of central charge 1.

\section{Steady-state density matrix in terms of local fields}\label{apploc}

Here we determine the steady state density matrix $\rho_{\mathrm{s}} = \exp -\widehat{W}$ where
\beq
	\h W = \frc{\beta_\rl + \beta_\rr}2 H
	+ \frc{\beta_\rl - \beta_\rr}2 \h (P_{\rm{1}}+\h Q).
\eeq
First note that 
\beq\label{hR}
	\h (P_{\rm{1}}+\h Q) = \int Dp\; {\rm sign}(p^1)\, E_p A^\dag_p A_p.
\eeq
The inversion of \eqref{modeexp} gives
\beq
	A_p = \int \mathrm{d}^dx\, \mathrm{e}^{-\mathrm{i}p\cdot x} \lt(E_p \phi(x) + \mathrm{i}\pi(x)\rt)
\eeq
which we insert into \eqref{hR}:
\beq\label{hr2}
	\h (P_{\rm{1}}+\h Q) = \int \mathrm{d}^dx \mathrm{d}^dy\,\Big[
	:\phi(x)\phi(y): Q'(x-y)\; + :\pi(x)\pi(y):Q''(x-y) \;+ :\phi(x)\pi(y): Q'''(x-y)
	\Big]
\eeq
where
\beqa
	Q'(x) &=& \int \frc{\mathrm{d}^dp}{2(2\pi)^d}\; {\rm sign}(p^1)\,
	E_p^2 \,\mathrm{e}^{\mathrm{i}p\cdot x} \n
	Q''(x) &=& \int \frc{\mathrm{d}^dp}{2(2\pi)^d}\; {\rm sign}(p^1)\,
	\mathrm{e}^{\mathrm{i}p\cdot x} \n
	Q'''(x) &=& i\int \frc{\mathrm{d}^dp}{(2\pi)^d}\; {\rm sign}(p^1)\,
	E_p \,\mathrm{e}^{\mathrm{i}p\cdot x}.
\eeqa
Clearly $Q'(x)$ and $Q''(x)$ are odd under $x\mapsto -x$, hence the first two terms in \eqref{hr2} vanish. On the other hand we can write
\beqa
	Q'''(x)&=& \mathrm{i}\int \frc{\mathrm{d}^{d-1}\t p}{(2\pi)^{d-1}}\, \mathrm{e}^{\mathrm{i}\t p\cdot \t x} \lt(
	\int \frc{\mathrm{d}p^1}{2\pi} p^1 \,\mathrm{e}^{\mathrm{i}p^1 x^1} +
	\int \frc{\mathrm{d}p^1}{2\pi} \; {\rm sign}(p^1)\,(E_p - |p^1|) \,\mathrm{e}^{\mathrm{i}p^1 x^1}\rt)
	\n
	&=& \p_1 \delta^{(d)}(x)+
	\mathrm{i}\int \frc{\mathrm{d}^{d-1}\t p}{(2\pi)^{d-1}} \,\mathrm{e}^{\mathrm{i}\t p\cdot \t x} \lt(
	\int \frc{\mathrm{d}p^1}{2\pi} \; {\rm sign}(p^1)\,(E_p - |p^1|) \,\mathrm{e}^{\mathrm{i}p^1 x^1}\rt)
	\no
\eeqa
where the tilde-variables represent transverse coordinates. The first term on the last line gives rise to the $P_1$ contribution in \eqref{hW}. For the second term, let us concentrate on the $p^1$-integral, which we write as
\beq
	\int_0^\infty
	\frc{\mathrm{d}p^1}{2\pi} \; (E_p - p^1) \,(\mathrm{e}^{\mathrm{i}p^1 x^1} - \mathrm{e}^{-\mathrm{i}p^1 x^1}).
\eeq
The large-$p^1$ behavior of $E_p - p^1$ is vanishing, hence the integral is convergent and we may shift the contours. Let us assume $x^1>0$; this is sufficient as the integral is odd under $x^1\mapsto -x^1$. Then for the terms proportional to $\mathrm{e}^{\pm \mathrm{i}p^1 x^1}$ we rotate the contours towards the positive/negative imaginary axis, $p^1 = \pm \mathrm{i}\ell$. The part proportional to $p^1$ cancels out, hence we obtain
\beq
	\mathrm{i}\int_0^\infty \frc{\mathrm{d}\ell}{2\pi} \lt(\sqrt{|\t p|^2 + m^2 -\ell^2+\mathrm{i}{\bf 0}}
	+ \sqrt{|\t p|^2 + m^2 -\ell^2-\mathrm{i}{\bf 0}}\rt)
	\mathrm{e}^{-x^1\ell}
	=
	\frc{\mathrm{i}}{\pi}\int_0^{E_{\t p}} \mathrm{d}\ell \,{\cal E}_{\ell,\t p}\,
	\mathrm{e}^{-x^1\ell}.
\eeq
This gives \eqref{kern}.

\section{Time evolution of fields} \label{appendixintegrals}

Here we compute the explicit form of the time evolved fields $\phi(x,t)$ and $\pi(x,t)$ in the $B$-representation.
For clarity, let us momentarily use the hat symbol in order to differentiate operators from fields. One can show that if $\h\phi(x,t)$ and $\h\pi(x,t)$ are operators obeying the canonical commutation relations \eqref{cr} (in any representation) and satisfying the equations of motion \eqref{eom}, then the functions
\beqa
	\phi(x,t) &=&
	-\mathrm{i}\int d^dy\,\lt([\h\phi(x,t),\h\pi(y)]\,\phi(y) - [\h\phi(x,t),\h\phi(y)]\,\pi(y)\rt) \n
	\pi(x,t) &=&
	-\mathrm{i}\int d^dy\,\lt([\h\pi(x,t),\h\pi(y)]\,\phi(y) - [\h\pi(x,t),\h\phi(y)]\,\pi(y)\rt),
	\label{te}
\eeqa
are solutions to the equations of motion \eqref{eom} with initial conditions $\phi(0,x) = \phi(x)$ and $\pi(0,x) = \pi(x)$. If we replace $\phi(y)$ and $\pi(y)$ by the $B$-representation \eqref{newmodeexp}, then by construction we have found operators $\h\phi(x,t)$ and $\h\pi(x,t)$ in the $B$-representation.

Using the $A$-representation, the operators $\h\phi(x,t)$ and $\h\pi(x,t)$ are given by \eqref{modeexpt}, and we may explicitly evaluate the commutators:
\beqa
	[\h\phi(x,t),\h\phi(y)] &=& -\mathrm{i}\int Dp\,2\sin (E_pt) \,\mathrm{e}^{\mathrm{i}p\cdot(x-y)} \n {}
	[\h\phi(x,t),\h\pi(y)] &=& \mathrm{i}\int Dp\,2E_p\,\cos(E_p t)\,\mathrm{e}^{\mathrm{i}p\cdot(x-y)} \n{}
	[\h\pi(x,t),\h\phi(y)] &=& -\mathrm{i}\int Dp\,2E_p\,\cos(E_p t)\,\mathrm{e}^{\mathrm{i}p\cdot(x-y)} \n{}
	[\h\pi(x,t),\h\pi(y)] &=& -\mathrm{i}\int Dp\,2E_p^2\,\sin (E_pt) \,\mathrm{e}^{\mathrm{i}p\cdot(x-y)}.
	\label{expl}
\eeqa
Putting \eqref{expl} and \eqref{newmodeexp} in \eqref{te} and evaluating the resulting integrals, we find
\beq
	\h\phi(x,t) = \int Dp\,B_p\,\int \frc{\mathrm{d}{\qq}}{2\pi}\,
	U(p^1,\qq)\,\mathrm{e}^{\mathrm{i}\qq x^1+\mathrm{i}\t p\cdot \t x} \lt( a^+_{{\qq},p} \mathrm{e}^{-\mathrm{i}E_{\qq,\t p}t}
	+ a^-_{\qq,p} \mathrm{e}^{\mathrm{i}E_{\qq,\t p}t}\rt) + \mathrm{h.c.}\label{itir}
\eeq
where
\beq
	a^\pm_{\qq,p} = \frc12\lt(1\pm \frc{E_p}{E_{\qq,\t p}}\rt),\quad
	E_{\qq,\t p} = \sqrt{|\t p|^2+m^2+\qq^2}
\eeq
and
\beqa
	U({\rm p},\qq)=
	\lt\{\ba{ll} \frc{\mathrm{i}}{\qq-{\rm p}+\mathrm{i}{\bf 0}} + \frc{\mathrm{i}}{\qq+{\rm p}+\mathrm{i}{\bf 0}} & \ ({\rm p}>0) \z
	 \frc{-\mathrm{i}}{\qq-{\rm p}-\mathrm{i}{\bf 0}} + \frc{-\mathrm{i}}{\qq+{\rm p}-\mathrm{i}{\bf 0}}  & \ ({\rm p}<0). \ea\rt.
	 \label{Upq}
\eeqa
Using \eqref{itir}, we may evaluate in the $B$-representation the operator $\h \pi(x,t)$ and any normal-ordered products of $\h \phi(x,t)$ and $\h \pi(x,t)$ and their derivatives. For the rest of this calculation we omit the hat symbol for field-operators.

Expression \eqref{itir} can be further simplified by contour deformations. Consider the $\qq$ integral in \eqref{itir}, omitting the factor $\mathrm{e}^{\mathrm{i}\t p\cdot \t x}$. By a change of variable, it can be written as
\beq\label{tinte}
	\int_0^\infty \frc{\mathrm{d}{\qq}}{2\pi}\,
	\lt(U(p^1,\qq)\,\mathrm{e}^{\mathrm{i}\qq x^1} + U(p^1,-\qq) \mathrm{e}^{-\mathrm{i}\qq x^1}\rt)
	\lt( a^+_{{\qq},p} \mathrm{e}^{-\mathrm{i}E_{\qq,\t p}t}
	+ a^-_{\qq,p} \mathrm{e}^{\mathrm{i}E_{\qq,\t p}t}\rt).
\eeq
We deform the $\qq$-contour by rotating either to $(0,\mathrm{i}\infty)$ or $(0,-\mathrm{i}\infty)$. The direction towards which we deform is determined by the values of $x^1$ and $t$. We note that at $\t p$ fixed, using the fact that $E_{\qq,\t p}\sim \qq$, the large-$\qq$ oscillating factors occur in four terms and are of the form $\mathrm{e}^{\mathrm{i}\qq x^1-\mathrm{i}\qq t}$, $\mathrm{e}^{\mathrm{i}\qq x^1+\mathrm{i}\qq t}$, $\mathrm{e}^{-\mathrm{i}\qq x^1-\mathrm{i}\qq t}$ and $\mathrm{e}^{-\mathrm{i}\qq x^1+\mathrm{i}\qq t}$. Hence, in order that no contribution at infinity be present upon contour deformation, we deform the $\qq$-contour towards the positive ($+$) or negative ($-$) imaginary direction as follows, respectively for each of the four terms (with $t>0$): 
\beq
	\ba{rccccc}
	&& \mathrm{e}^{\mathrm{i}\qq x^1-\mathrm{i}\qq t} & \mathrm{e}^{\mathrm{i}\qq x^1+\mathrm{i}\qq t} & \mathrm{e}^{-\mathrm{i}\qq x^1-\mathrm{i}\qq t} &
	\mathrm{e}^{-\mathrm{i}\qq x^1+\mathrm{i}\qq t}\\
	x^1>t &:& + &+& -& - \\
	-t<x^1<t &:& -&+&-&+ \\
	x^1<-t &:& -&-&+&+ \ea
	\label{defo} 
\eeq

Upon deformation, singularities of the function $U(p^1,\qq)$ are crossed. These are simple poles at $\qq = \pm p^1$. Because of the imaginary shift $\pm \mathrm{i}{\bf 0}$ in \eqref{Upq}, they are crossed only if $p^1<0$ when deforming the $\qq$-contour towards the positive imaginary direction, and only if $p^1>0$ when deforming the $\qq$-contour towards the negative imaginary direction. Remembering that $\qq>0$, only one of the two poles is crossed in any case. In both cases of the sign in $\qq = \pm p^1$, at the position of these poles we have $E_{q,\t p} = E_p$. Hence, at the position of these poles, $a^+_{\qq,p} = 1$ and $a^-_{\qq,p}=0$. This means that in order to evaluate the associated residues, it is sufficient to consider only the terms with large-$\qq$ factors $\mathrm{e}^{\mathrm{i}\qq x^1-\mathrm{i}\qq t}$ and $\mathrm{e}^{-\mathrm{i}\qq x^1-\mathrm{i}\qq t}$.

In the case $x^1>t$, we shift in the directions ($+$) and ($-$) for the terms containing $\mathrm{e}^{\mathrm{i}\qq x^1-\mathrm{i}\qq t}$ and $\mathrm{e}^{-\mathrm{i}\qq x^1-\mathrm{i}\qq t}$, respectively. In the first shift, we cross the pole of $U(p^1,\qq)$ at $\qq = -p^1$, and in the second, we cross the pole of $U(p^1,-\qq)$ at $-\qq = p^1$, in both cases only if $p^1<0$. One can do a similar analysis for $x^1<-t$. The result, including the hermitian conjugate, is 
\beq\label{xppt}
	\phi(x,t) \stackrel{|x^1|>t}= \int Dp\,(B_p \mathrm{e}^{\mathrm{i}\t p \cdot \t x - \mathrm{i} E_{p}t} + B_p^\dag \mathrm{e}^{-\mathrm{i}\t p \cdot \t x+ \mathrm{i} E_{p}t} )\,2\cos (p_1x_1) \Theta(-p_1x_1) +
	\mbox{integral contribution}
\eeq
where the integral contribution, calculated below, comes from the shifted integral itself. The terms explicitly written on right-hand side are, according to the first equation of \eqref{newmodeexp}, equal to $\mathrm{e}^{\mathrm{i}H_0t}\phi(x) \mathrm{e}^{-\mathrm{i}H_0t}$ where
\beq\label{H0}
	H_0 = H_{\rm L}+H_{\rm R}.
\eeq

In the case $-t<x^1<t$, we shift in the direction ($-$) for both terms containing $ \mathrm{e}^{\mathrm{ i}\qq x^1-\mathrm{i}\qq t}$ and $\mathrm{e}^{-\mathrm{i}\qq x^1-\mathrm{i}\qq t}$. In the first shift we cross the pole of $U(p^1,\qq)$ at $\qq=p^1$ if $p^1>0$, and in the second, we cross the pole of $U(p^1,-\qq)$ at $-\qq=p^1$ if $p^1<0$. Together, we obtain
\beq
	\phi(x,t)\stackrel{|x^1|<t}= \int D p \,(B_p \mathrm{e}^{-\mathrm{i}E_pt+\mathrm{i}p\cdot x} + B^\dag_p \mathrm{e}^{\mathrm{i}E_pt-\mathrm{i}p\cdot x}) + \mbox{integral contribution}
	\label{ppo}
\eeq
where again the integral contribution coems from the shifted integral itself. The explicit terms on the right-hand side have the structure of \eqref{modeexpt}, except for the replacement $A_p,A^\dag_p\mapsto B_p,B^\dag_p$. That is, recalling the scattering isomorphism $S$ \eqref{Smat}, we may write them as $S(\phi(x,t))$.

Finally, we calculate the integral contributions, from the shifted ${\rm q}$-integral. After performing the deformations the $\qq$-integral \eqref{tinte} runs, for the various terms, between $0$ and $\pm \mathrm{i}\infty$. In order to assess the result, we have to separate the region $|\qq|>E_{\t p}:=\sqrt{|\t p|^2+m^2}$ from the region $|\qq|<E_{\t p}$, because there are branch points at $\qq = \pm \mathrm{i} E_{\t p}$ with branch cuts going towards $\pm \mathrm{i}\infty$. After deformation, we change the $\qq$ variable to $\qq = \pm \mathrm{i} w$ and we are left with integrals $\int_0^{\pm \mathrm{i}\infty} \mathrm{d}q = \pm \mathrm{i}\int_0^\infty \mathrm{d}w$. In the case $w>E_{\t p}$, we may use $E_{\pm \mathrm{i}w,\t p} = \pm \mathrm{i} \sqrt{w^2 - |\t p|^2-m^2}$, which implies $a^\pm_{\mathrm{i}w,p} = a^\mp_{-\mathrm{i}w,p}$. On the other hand, in the case $w<E_{\t p}$, we have $E_{\pm \mathrm{i}w,\t p} = \sqrt{|\t p|^2+m^2-w^2}$ and $a^\pm_{ \mathrm{i}w,p} = a^\pm_{-\mathrm{i}w,p}$.

Putting these rules together, in the case $|x^1|>t$, a straightforward calculation shows that both in the regions $w>E_{\t p}$ and $0<w<E_{\t p}$, the integrand in \eqref{tinte}, arising after the deformation $+,+,-,-$ or $-,-,+,+$ (as per \eqref{defo}), is exactly zero. That is, the integral contribution is exactly zero in \eqref{xppt}.

In the case $|x^1|<t$, in the region $w>E_{\t p}$ the integrand \eqref{tinte} also gives zero after the deformation $-,+,-,+$, but it is non-zero in the region $0<w<E_{\t p}$. This contribution is (with $p=(p^1,\t p)$)
\beq
	G(x^1,t;p) := \frc{\mathrm{i}\,{\rm sign}(p^1)}\pi \int_0^{E_{\t p}} \mathrm{d}w\,\frc{w\sinh(wx^1)}{w^2+(p^1)^2}\lt(
	\lt(1+\frc{E_p}{{\cal E}_{w,\t p}}\rt) \mathrm{e}^{-\mathrm{i}{\cal E}_{w,\t p} t}
	-
	\lt(1-\frc{E_p}{{\cal E}_{w,\t p}}\rt) \mathrm{e}^{\mathrm{i}{\cal E}_{w,\t p} t}
	\rt)
\eeq
where
\beq
	{\cal E}_{w,\t p} = \sqrt{|\t p|^2+m^2-w^2}.
\eeq
Changing variable to $u = {\cal E}_{w,\t p}$, this simplifies to \eqref{corrF}

Recall that the above calculation was performed with initial (before-quench) free condition on $\phi(x)$ at $x^1=0$. A similar calculation may be done for initial fixed condition at $x^1=0$. The factor $\cos(p^1 x^1)$ is replaced by $i\sin (p^1 x^1)$ in \eqref{newmodeexp}. The result is as above, but with the correction factor \eqref{corrFf} instead of that given by \eqref{corrF}. Technically, this may be obtained by observing that one can go from fixed to free condition by formally applying the operator $(\mathrm{i}p^1)^{-1} \p/\p x^1$ inside the $p$-integral (see the expression \eqref{newmodeexp}, with $\cos(p^1 x^1)$ is replaced by $i\sin (p^1 x^1)$); the sine function guarantees that the delta-function coming from the factor $\Theta(-p^1x^1)$ does not contribute.

Finally, let us discuss the convergence properties of expressions obtained from \eqref{result}. Consider evaluating the average of a {\em local} operator $\Or(x)$ at $x$, formed by normal-ordered products of $\phi(x)$, $\pi(x)$ and their derivatives. One uses \eqref{result} and \eqref{trbb} along with Wick's theorem. Integrating over momenta, we see that the integrands will be suppressed by factors $e^{-|p| \beta_{\rl,\rr}}$ at large momenta for every Wick contraction, due to the denominator in the second equation of \eqref{trbb} (the one that is relevant for normal-ordered operators). Contractions that involve two $G$-corrections have two factors $G(x^1,t;p)$, integrated over $p$. Due to the $\sinh$ (resp.~$\cosh$) factor in the integrand in \eqref{corrF} (resp.~\eqref{corrFf}), the large-$|\t p|$ behaviour of the integrand along the integration path has a factor $e^{2|\t p|\,|x^1|}$ (for $u$ away from the integration limits). Hence, this shows convergence of the resulting integral only for $|x^1|<2\,{\rm min}(\beta_{\rl,\rr})$. For values of $x^1$ beyond this region, one may modify the integral representation to show convergence beyond this region. A convenient way is to shift the $u$ contour towards the positive imaginary direction. The result, which we report here for the initially free case, is:
\beqa
	\lefteqn{G(x^1,t;p)} &&\n &=& -\frc{{\rm sign}(p^1)}\pi
	\int_0^\infty \mathrm{d}v\,e^{-vt} \lt(
	\frc{\sinh(\sqrt{v^2+2ivE_{\t p}}\,x^1)}{E_p+\mathrm{i}v-E_{\t p}}
	\mathrm{e}^{-\mathrm{i}E_{\t p}\,t}-
	\frc{\sinh(\sqrt{v^2-2\mathrm{i}vE_{\t p}}\,x^1)}{E_p+\mathrm{i}v+E_{\t p}}
	\mathrm{e}^{\mathrm{i}E_{\t p}\,t}\rt) \n &&
	\quad \mbox{(initially free at $x^1=0$)}
\eeqa
The integrand now diverges much more slowly at large $|\t p|$, which guarantees convergence. We use such representations in the next subsection in order to perform the large-time asymptotic analysis of the energy current average.

\section{Asymptotic time evolution of the energy current}
\label{appt01}

We consider the large-time asymptotic of the average $\bra T^{01}(t,0)\ket$ with fixed boundary conditions at $d=1$, evaluated using \eqref{result} and \eqref{tmunu}.

Performing the trace and then the integration over the angles in the
transverse direction, we find, in generic dimensions $d$,
\beqa\label{avercud} \bra T^{01}(x,t)\ket &\stackrel{|x^1|<t}=& \bra
T^{01}\ket_{\rm s} + \frc1{2^{d}\pi^{\frc{d+1}2}
  \Gamma\lt(\frc{d-1}2\rt)} \int_{-\infty}^\infty \mathrm{d}p^1
\int_0^\infty \mathrm{d}\t p\, \frc{{\t p}^{d-2} }{E_p\,
  (\mathrm{e}^{W(p)}-1)} \times \\ && \times\; \lt[ 2p^1 {\rm
    Im}\lt(\mathrm{e}^{\mathrm{i}p^1x^1-\mathrm{i}E_p t} \p_0 \b G\rt)
  +2E_p {\rm Im}\lt(\mathrm{e}^{-\mathrm{i}p^1x^1+\mathrm{i}E_p t}
  \p_1 G\rt) - 2 {\rm Re}\lt(\p_0 G \p_1 \b G\rt)\rt].\no \eeqa For simplicity, we
will restrict ourselves to $d=1$. For $d=1$ the derivation, without the angular
integration in the transverse direction (and omitting the upper-1
space index), gives \beqa\label{avercu} \bra T^{01}(x,t)\ket
&\stackrel{|x|<t,\,d=1}=& \bra T^{01}\ket_{\rm s} +
\int_{-\infty}^\infty \frc{\mathrm{d}p}{4\pi E_p\,
  (\mathrm{e}^{W(p)}-1)} \times \\ && \times\; \lt[ 2p {\rm
    Im}\lt(\mathrm{e}^{\mathrm{i}px-\mathrm{i}E_p t} \p_0 \b G\rt)
  +2E_p {\rm Im}\lt(\mathrm{e}^{-\mathrm{i}px+\mathrm{i}E_p t} \p_1
  G\rt) - 2 {\rm Re}\lt(\p_0 G \p_1 \b G\rt)\rt].\no \eeqa We use
fixed initial boundary conditions, for which the function $G$ is \beq
G(x^1,t;p) = -\frc{|p^1|}\pi \int_{-E_{\t p}}^{E_{\t p}} \mathrm{d}u
\, \frc{ \cosh\lt({\cal E}_{u,\t p}\,x^1\rt)}{(E_p+u){\cal E}_{u,\t
    p}} \mathrm{e}^{\mathrm{i}ut}.  \eeq Again, for $d=1$ this
specializes to \beq\label{G1} G(x,t;p) = -\frc{|p|}\pi \int_{-m}^{m}
\mathrm{d}u \, \frc{
  \cosh\lt(\sqrt{m^2-u^2}\,x\rt)}{(E_p+u)\sqrt{m^2-u^2}}
\mathrm{e}^{\mathrm{i}ut}.  \eeq

We have \beq \p_1 G = -\frc{|p|}\pi \int_{-m}^{m} \mathrm{d}u \, \frc{
  \sinh\lt(\sqrt{m^2-u^2}\,x\rt)}{(E_p+u)} \mathrm{e}^{\mathrm{i}ut}.
\eeq Specializing at $x=0$, this is zero. Hence, in the expression for
the current \eqref{avercu}, only the first term remains and we have
\beqa\label{avercu2} \bra T^{01}(t,0)\ket = \bra T^{01}\ket_{\rm s}
+ \int_{-\infty}^\infty \frc{\mathrm{d}p}{4\pi E_p\,
  (\mathrm{e}^{W(p)}-1)} 2p\, {\rm Im}\lt(\mathrm{e}^{-\mathrm{i}E_p
  t} \p_0 \b G(0,t;p)\rt) \eeqa where \beq\label{G2} G(0,t;p) =
-\frc{|p|}\pi \int_{-m}^{m} \mathrm{d}u \, \frc{
  \mathrm{e}^{\mathrm{i}ut}}{(E_p+u)\sqrt{m^2-u^2}}.  \eeq We are
interested in the correction \beq \delta(\beta_\rl,\beta_\rr) = \int_{-\infty}^\infty
\frc{\mathrm{d}p}{4\pi E_p\, (\mathrm{e}^{W(p)}-1)} 2p\, {\rm
  Im}\lt(\mathrm{e}^{-\mathrm{i}E_p t} \p_0 \b G(0,t;p)\rt).  \eeq

We now recast \eqref{G2} in a form where the asymptotic analysis can
be made accurately. We deform the contour towards the positive
imaginary direction, where it vanishes at positive imaginary
infinity. There are two contributions, one on the line with real part
$-m$ the other with real part $m$. For the first we change variable to
$u=-m+\mathrm{i}v$ and for the other, $u=m+\mathrm{i}v$, with $v$ from
$0$ to $\infty$ in the first instance, and from $\infty$ to 0 in the
second. This gives \beq G(0,t;p) = -\frc{\mathrm{i}\,|p|}\pi
\int_{0}^{\infty} \mathrm{d}v\,\mathrm{e}^{-vt}\lt(
\frc{\mathrm{e}^{-\mathrm{i}mt}}{(E_p-m+\mathrm{i}v)\sqrt{v^2+2\mathrm{i}mv}}
-
\frc{\mathrm{e}^{\mathrm{i}mt}}{(E_p+m+\mathrm{i}v)\sqrt{v^2-2\mathrm{i}mv}}
\rt) \eeq and \beqa \p_0 \b G(0,t;p) &=& \frc{i\,|p|}\pi
\int_{0}^{\infty} \mathrm{d}v\,\mathrm{e}^{-vt}\lt(
\frc{(\mathrm{i}m-v)\mathrm{e}^{\mathrm{i}mt}}{(E_p-m-\mathrm{i}v)\sqrt{v^2-2\mathrm{i}mv}}
+
\frc{(\mathrm{i}m+v)\mathrm{e}^{-\mathrm{i}mt}}{(E_p+m-\mathrm{i}v)\sqrt{v^2+2\mathrm{i}mv}}
\rt)\n &=& g_+ + g_- \eeqa where \beq g_\pm = \frc{\mathrm{i}\,|p|}\pi
\int_{0}^{\infty} \mathrm{d}v\,\mathrm{e}^{-vt} \frc{( \mathrm{i}m\mp
  v)\mathrm{e}^{\pm \mathrm{i}mt}}{(E_p\mp m-\mathrm{i}v)\sqrt{v^2\mp
    2\mathrm{i}mv}}.  \eeq

We consider the integral
\beq
	I(\beta,t) = \int_0^\infty \frc{\mathrm{d}p\,p\,\mathrm{e}^{-\mathrm{i}E_p t }\,\p_0 \b G(0,t;p)}{
	E_p (\mathrm{e}^{\beta E_p}-1)} = I_+ + I_-
\eeq
where
\beq
	I_\pm = \int_0^\infty \frc{\mathrm{d}p\,p\,\mathrm{e}^{-\mathrm{i}E_p t }\,g_\pm}{
	E_p (\mathrm{e}^{\beta E_p}-1)}.
\eeq
The correction to the stationary value is given by
\beq
	\delta(\beta_\rl,\beta_\rr) = \frc1{2\pi} {\rm Im}\lt(I(\beta_l,t)-I(\beta_r,t)\rt).
\eeq
Changing variables we have
\beqa
	I_\pm &=& \frc{\mathrm{i}}{\pi}
	\int_m^\infty \mathrm{d}E\,\frc{\mathrm{e}^{-\mathrm{i}(E\mp m)t}}{\mathrm{e}^{\beta E}-1}
	\int_0^\infty \mathrm{d}v\,\frc{\mathrm{e}^{-vt}\, (\mathrm{i}m\mp v)\,\sqrt{E^2-m^2}}{(E\mp m-\mathrm{i}v)
	\sqrt{v^2\mp 2\mathrm{i}mv}}.
\eeqa

Let us concentrate on $I_+$. We find
\beqa
	I_+&=& \frc{\mathrm{i}}{\pi}
	\int_m^\infty \mathrm{d}E\,\frc{\mathrm{e}^{-\mathrm{i}(E- m)t}}{\mathrm{e}^{\beta E}-1}
	\int_0^\infty \mathrm{d}v\,\frc{\mathrm{e}^{-vt}\, (\mathrm{i}m- v)\,\sqrt{E^2-m^2}}{(E- m-\mathrm{i}v)
	\sqrt{v^2- 2\mathrm{i}mv}} \n
	&=& \frc{\mathrm{i}}{\pi}
	\int_0^\infty \mathrm{d}E\,\frc{\mathrm{e}^{-\mathrm{i}Et}}{\mathrm{e}^{\beta (E+m)}-1}
	\int_0^\infty \mathrm{d}v\,\frc{\mathrm{e}^{-vt}\, (\mathrm{i}m- v)\,\sqrt{E} \sqrt{E+2m}}{(E-\mathrm{i}v)
	\sqrt{v^2- 2\mathrm{i}mv}}. \no
\eeqa
In terms of the variable $E$, there are singularities at $0$, $-2m$, $-m+2\pi \mathrm{i} n/\beta\;(n\in\Z)$ and $iv$. Hence there are no singularities in the lower-right quadrant, and we can deform the contour towards the negative imaginary direction. With the replacement $E\mapsto -iE$, this gives
\beqa
I_+	&=& \frc{\mathrm{i}(-\mathrm{i})^{\frc12}}{\pi}
	\int_0^\infty \mathrm{d}E\,\frc{\mathrm{e}^{-Et}}{\mathrm{e}^{\beta (m-\mathrm{i}E)}-1}
	\int_0^\infty \mathrm{d}v\,\frc{\mathrm{e}^{-vt}\, (\mathrm{i}m- v)\,\sqrt{E} \sqrt{2m-\mathrm{i}E}}{(E+v)
	\sqrt{v^2- 2\mathrm{i}mv}}.
\eeqa
Since we have real exponentials $\mathrm{e}^{-Et}$ and $\mathrm{e}^{-vt}$, the asymptotic large-$t$ expansion is obtained by expanding the integrand around $E=0$ and $v=0$. We do that to first order. We have
\beqa
	\sqrt{2m-\mathrm{i}E} &\approx& \sqrt{2m} \lt(1-\frc{\mathrm{i}E}{4m}\rt) \\
	\frc1{\mathrm{e}^{\beta(m-\mathrm{i}E)}-1} &\approx&
	\frc1{\mathrm{e}^{\beta m}(1-\mathrm{i}\beta E)-1} \n
	&=&
	\frc1{(\mathrm{e}^{\beta m}-1)\lt(1-\frc{\mathrm{i}\beta E}{1-\mathrm{e}^{-\beta m}}\rt)} \n
	&\approx&
	\frc1{\mathrm{e}^{\beta m}-1}\lt(1+\frc{\mathrm{i}\beta E}{1-\mathrm{e}^{-\beta m}}\rt) \\
	\frc1{\sqrt{v^2- 2\mathrm{i}mv}} &\approx& \frc1{\sqrt{-2\mathrm{i}m} \sqrt{v}}\lt(
	1+\frc{v}{4\mathrm{i}m}\rt)\\
	\mathrm{i}m-v &=& im\lt(1-\frc{v}{\mathrm{i}m}\rt).
\eeqa
This gives
\beq
	I_+ \approx -\frc{m}{\pi}\frc1{\mathrm{e}^{\beta m}-1} \int_0^\infty \mathrm{d}E\,\mathrm{d}v\,
	\frc{e^{-(E+v)t}}{E+v} \sqrt{\frc{E}v} \lt(
	1 -\frc{\mathrm{i}E}{4m} + \frc{\mathrm{i}\beta E}{1-\mathrm{e}^{-\beta m}} + \frc{v}{4\mathrm{i}m}- \frc{v}{\mathrm{i}m}
	\rt).
\eeq
We now evaluate the following integrals:
\beqa
	\int_0^\infty \mathrm{d}E\,\mathrm{d}v\,\frc{\mathrm{e}^{-(E+v)t}}{E+v} \sqrt{\frc{E}v} \,E^j\,v^k
	&=& \int_t^\infty \mathrm{d}s\,
	\int_0^\infty \mathrm{d}E\,\mathrm{d}v\,\mathrm{e}^{-(E+v)s} \sqrt{\frc{E}v} \,E^j\,v^k \n
	&=& \int_t^\infty \mathrm{d}s\,
	s^{-2-j-k}\, \Gamma(j+3/2)\Gamma(k+1/2) \n
	&=& \frc{t^{-1-j-k}}{1+j+k} \, \Gamma(j+3/2)\Gamma(k+1/2)\\
	&=& \lt\{\ba{ll}
	\frc{\pi}{2t} & (j=k=0) \z
	\frc{3\pi}{8t^2} & (j=1,\;k=0) \z
	\frc{\pi}{8t^2} & (j=0,\;k=1).
	\ea\rt.
\eeqa
Putting these together,
\beqa
	I_+ &\approx& -\frc{m}{2t}\,\frc1{\mathrm{e}^{\beta m}-1} \lt(
	1 -\frc{3\mathrm{i}}{16mt}+\frc3{4t}\,
	\frc{\mathrm{i}\beta }{1-\mathrm{e}^{-\beta m}} - \frc{3}{16\mathrm{i}mt}
	\rt)\n
	&=& -\frc{m}{2t}\,\frc1{\mathrm{e}^{\beta m}-1} \lt(
	1 +\frc3{4t}\,
	\frc{\mathrm{i}\beta }{1-\mathrm{e}^{-\beta m}}
	\rt).
\eeqa
Hence this contribution is
\beq
	\frc1{2\pi} {\rm Im}(I_+) = -\frc{3m\beta}{64\pi \sinh^2(\beta m/2)} t^{-2}
	+O(t^{-3}).
\eeq

Let us next concentrate on $I_-$. We have, following similar lines,
\beqa
	I_- &=& \frc{\mathrm{i}}{\pi}
	\int_m^\infty \mathrm{d}E\,\frc{\mathrm{e}^{-\mathrm{i}(E+ m)t}}{\mathrm{e}^{\beta E}-1}
	\int_0^\infty \mathrm{d}v\,\frc{\mathrm{e}^{-vt}\, (\mathrm{i}m+ v)\,\sqrt{E^2-m^2}}{(E+ m-\mathrm{i}v)
	\sqrt{v^2+ 2\mathrm{i}mv}} \n
	&=& \frc{\mathrm{i}}{\pi}
	\int_0^\infty \mathrm{d}E\,\frc{\mathrm{e}^{-\mathrm{i}(E+ 2m)t}}{\mathrm{e}^{\beta (E+m)}-1}
	\int_0^\infty \mathrm{d}v\,\frc{\mathrm{e}^{-vt}\, (\mathrm{i}m+ v)\,\sqrt{E} \sqrt{E+2m}}{(E+ 2m-\mathrm{i}v)
	\sqrt{v^2+ 2\mathrm{i}mv}} \n
	&=& \frc{\mathrm{i}(-\mathrm{i})^{\frc12}\mathrm{e}^{-2\mathrm{i}mt}}{\pi}
	\int_0^\infty \mathrm{d}E\,\frc{\mathrm{e}^{-Et}}{\mathrm{e}^{\beta (m-\mathrm{i}E)}-1}
	\int_0^\infty \mathrm{d}v\,\frc{\mathrm{e}^{-vt}\, (\mathrm{i}m+ v)\,\sqrt{E} \sqrt{2m-\mathrm{i}E}}{(E+v+ 2\mathrm{i}m)
	\sqrt{v^2+ 2\mathrm{i}mv}}.
\eeqa
Now it is sufficient to expand to zeroth order, obtaining
\beqa
I_-	&\approx& \frc{\mathrm{i}(-\mathrm{i})^{\frc12}\mathrm{e}^{-2\mathrm{i}mt}}{2\pi(\mathrm{e}^{\beta m}-1)}
	\int_0^\infty \mathrm{d}E\,\mathrm{e}^{-Et}
	\int_0^\infty \mathrm{d}v\,\frc{\mathrm{e}^{-vt}\, \sqrt{E} \sqrt{2m}}{
	\sqrt{2\mathrm{i}mv}} \n
	&=& \frc{\mathrm{e}^{-2\mathrm{i}mt}}{2\pi(\mathrm{e}^{\beta m}-1)}
	\int_0^\infty \mathrm{d}E\,\mathrm{d}v\,\mathrm{e}^{-(E+v)t}
	\sqrt{\frc{E}v} \n
	&=& \frc{\mathrm{e}^{-2\mathrm{i}mt}}{2\pi(\mathrm{e}^{\beta m}-1)}
	t^{-2} \Gamma(3/2)\Gamma(1/2) \n
	&=& \frc{\mathrm{e}^{-2\mathrm{i}mt}}{4(e^{\beta m}-1)}
	t^{-2}.
\eeqa
Hence the contribution is
\beq
	\frc1{2\pi}{\rm Im}(I_-) = -\frc{\sin(2mt)}{8\pi (\mathrm{e}^{\beta m}-1)} t^{-2}
	+O(t^{-3}).
\eeq

Putting the above results together, we find $\delta(\beta_\rl,\beta_\rr) = \delta(\beta_{\mathrm{L}}) - \delta(\beta_{\mathrm{R}})$ where $\delta(\beta)$ is given in \eqref{delta}. We note that the correction has an oscillatory part, and that it is always negative (for all $\beta_{\mathrm{L,R}} m>0$ and $\beta_{\mathrm{L}}<\beta_{\mathrm{R}}$), with $\delta(\infty)=0$.

\section{Numerical simulations}
\label{App:Numerics}

An independent check of our analytical calculations are provided by a
first principles numerical simulation of the evolution of the free
field theory.  We will solve this problem for the case $d=1$, for a
massive boson.  The higher dimensional cases may be solved by this
method as well, as the transverse momenta act as an effective
mass $m_{\mathrm{eff}}^2\equiv\sqrt{\tilde{p}^2+m^2}$, from the point
of view of any observable dependent only on $x$ and $t$.  The simplest
way to regulate this theory numerically is to place the system in a
large, but finite box, of length $\mathcal{L}$.  Let us denote
$0<X<\mathcal{L}$ with the spatial coordinate of the box -- in the
continuum theory, $x = X-\mathcal{L}/2$.  Choosing boundary conditions
$\phi(X=0) = \phi(X=\mathcal{L})=0$, we may write the (classical)
field \begin{equation} \phi(x,t) = \sum_{n=1}^\infty a_n(t)
  \sin\left(\frac{n\pi X}{\mathcal{L}}\right),
\end{equation}

As in the main text, it will suffice to compute
solutions of the classical equations of motion, to solve the full
quantum evolution for a free theory.  In particular, we need to
time-evolve the finite-box equivalent of the half-modes $\cos(p\cdot
x) \Theta(-p^1x)$.  These modes are characterized by an integer $m$,
which we take to be odd if we wish for, in the continuum theory,
$\phi$ to satisfy free boundary conditions at the interface; $m$ is
even if the boundary conditions are fixed.  Without loss of
generality, let us consider modes propagating from left to right.
Since this is a free theory, the appropriate Green's functions are
given by \begin{equation} \mathcal{G}_l(t=0) = \sum_{n=1}^\infty
  \mathcal{A}_{l,n}(t) \sin \frac{n\pi X}{\mathcal{L}}
\end{equation}where \begin{equation}
\mathcal{A}_{l,n}(t) \equiv \left\lbrace \begin{array}{ll}
  \dfrac{\sqrt{2}}{\pi} \left(\dfrac{\sin((l-n)\pi/2)}{l-n} -
  \dfrac{\sin((l+n)\pi/2)}{l+n} \right)\left(\cos(E_n t) -
  \mathrm{i}\dfrac{E_l}{E_n}\sin(E_nt)\right) &\ l \ne n
  \\ \dfrac{\mathrm{e}^{-\mathrm{i}E_l t}}{\sqrt{2}} &\ l =
  n \end{array}\right.
\end{equation}with \begin{equation}
E_n \equiv \sqrt{m^2+ \frac{n^2\pi^2}{\mathcal{L}^2}}.
\end{equation}

These Green's functions are straightforwardly used to evaluate
correlation functions.  For example, let us consider the case where
$d=1$, $T_{\mathrm{L}}=T$, and $T_{\mathrm{R}}=0$.
Then \begin{equation} \langle T^{01}(X,t)\rangle = \sum_l \frac{2}{\mathcal{L}}
  \frac{1}{\mathrm{e}^{E_l/T}-1} \left[-\mathrm{Re}\left(\partial_x
    \mathcal{G}_l \partial_t \mathcal{G}_l\right)\right],
\end{equation}
where the sum over $m$ is only over odd positive integers (free
boundary conditions) or even positive integers (fixed boundary
conditions).

In practice, we must truncate the sum over integers to a finite
number.  We have typically chosen $T,m\sim \mathcal{O}(1)$,
$\mathcal{L}=200$ and truncated the sum to include $m \le 800$, which
is adequate to find convergence.

\section{Correlation functions in the steady state}
\label{appcorr}

Here we compute the algebraic decay of the $\langle \phi(x)\pi(0)\rangle$ correlation function in $d=1$ massive Klein-Gordon theory at late times after the local quench.
In order to show \eqref{phipis}, we use the trace relations \eqref{traa} along with \eqref{modeexp}. We find, in $d=1$,
\beqa
	\bra\phi(x)\pi(0)\ket_{\rm s} &=&
	i\int_{-\infty}^\infty \frc{\mathrm{d}p}{4\pi}\lt(
	\frc{\mathrm{e}^{\mathrm{i}px^1}}{1-\mathrm{e}^{-W(p)}} -\frc{\mathrm{e}^{-\mathrm{i}px^1}}{\mathrm{e}^{W(p)}-1}\rt) \n
	&=& 
	i\int_{0}^\infty \frc{dp}{4\pi}\lt(
	\frc{\mathrm{e}^{\mathrm{i}px^1}}{1-\mathrm{e}^{-\beta_\rl E_p}} -\frc{\mathrm{e}^{-\mathrm{i}px^1}}{\mathrm{e}^{\beta_{\rl}E_p}-1}\rt) + ((\beta_\rl,x^1)\mapsto (\beta_\rr,-x^1)).
\eeqa
The asymptotics of the integral can be evaluated by contour deformation. Let us assume $x^1>0$. Then:
\beqa
	\lefteqn{\mathrm{i}\int_{0}^\infty \frc{\mathrm{d}p}{4\pi}\lt(
	\frc{\mathrm{e}^{\mathrm{i}px^1}}{1-\mathrm{e}^{-\beta_\rl E_p}} -\frc{\mathrm{e}^{-\mathrm{i}px^1}}{\mathrm{e}^{\beta_{\rl}E_p}-1}\rt)} &&\n &=& 
	-\int_0^m \frc{d\ell}{4\pi}e^{-\ell x^1}\lt(
	\frc{1}{1-\mathrm{e}^{-\beta_\rl \sqrt{m^2-\ell^2}}} +\frc{1}{\mathrm{e}^{\beta_{\rl}\sqrt{m^2-\ell^2}}-1}\rt) +O(\mathrm{e}^{-mx^1}) \n
&\sim& 
	-\int_0^\infty \frc{\mathrm{d}\ell}{4\pi}\mathrm{e}^{-\ell x^1} \coth\frc{\beta_\rl m}2 \n
&=& 
	-\frc1{4\pi x^1} \coth\frc{\beta_\rl m}2
\eeqa
Hence we find \eqref{phipis}. A similar analysis for $\bra\phi(x)\phi(0)\ket_{\rm s}$ gives an {\em exponential decay} instead of an algebraic decay. Indeed, we have
\beqa\label{phiphi}
	\bra\phi(x)\phi(0)\ket_{\rm s} 
	&=& 
	\int_{0}^\infty \frc{\mathrm{d}p}{4\pi E_p}\lt(
	\frc{\mathrm{e}^{\mathrm{i}px^1}}{1-\mathrm{e}^{-\beta_\rl E_p}} +\frc{\mathrm{e}^{-\mathrm{i}px^1}}{\mathrm{e}^{\beta_{\rl}E_p}-1}\rt) + ((\beta_\rl,x^1)\mapsto (\beta_\rr,-x^1))
\eeqa
and
\beqa
	\lefteqn{\int_{0}^\infty \frc{\mathrm{d}p}{4\pi E_p}\lt(
	\frc{\mathrm{e}^{\mathrm{i}px^1}}{1-\mathrm{e}^{-\beta_\rl E_p}} +\frc{\mathrm{e}^{-\mathrm{i}px^1}}{\mathrm{e}^{\beta_{\rl}E_p}-1}\rt)} &&\n
	&=& 
	i\int_0^m \frc{\mathrm{d}\ell}{4\pi\sqrt{m^2-\ell^2}}\mathrm{e}^{-\ell x^1}\lt(
	\frc{1}{1-e^{-\beta_\rl \sqrt{m^2-\ell^2}}} -\frc{1}{e^{\beta_{\rl}\sqrt{m^2-\ell^2}}-1}\rt) +O(\mathrm{e}^{-mx^1}) \n
	&=& 
	i\int_0^m \frc{\mathrm{d}\ell}{4\pi\sqrt{m^2-\ell^2}}\mathrm{e}^{-\ell x^1} +O(\mathrm{e}^{-mx^1})\n
	&=& 
	\sum_{k=0}^\infty \frc{a_k}{x^{2k+1}} +O(\mathrm{e}^{-mx^1}) \no
\eeqa
where the constants $a_k$ are obtained from expanding the integrand in $\ell^2$ in the penultimate line. Putting this into \eqref{phiphi}, the algebraic part cancels out, thus showing exponential decay of the correlation function.

\section{Dimensional reduction} \label{app1d}

Dimensional reduction through integration over ``perpendicular''
coordinates is a phenomenon that has been observed in various
situations, perhaps the most prominent being the reduction of
electronic leads from three to one dimension in the context of
impurity models. Here we present the general theory for the
Klein--Gordon model.

Let $1\leq D<d$ and consider the following fields, which are integration over a $(d-D)$-dimensional subspace of the fundamental Klein--Gordon fields:
\beq\label{mapd}
	\Phi({\rm x}):=\frc1{L^{\frc{D-d}2}}\int_{-\frc L2}^{\frc L2} \mathrm{d}^{d-D}\t x\,\phi(x),\quad
	\Pi({\rm x}):=\frc1{L^{\frc{D-d}2}}\int_{-\frc L2}^{\frc L2} \mathrm{d}^{d-D}\t x\,\pi(x)
\eeq
where ${\rm x}=(x^1,\ldots,x^{D})$ and $ \t x = (x^{D+1},\ldots, x^d)$. Consider also their time-evolution $\Phi(t,{\rm x}) = \mathrm{e}^{\mathrm{i}Ht} \Phi({\rm x})\mathrm{e}^{-\mathrm{i}Ht}$, $\Pi(t,{\rm x}) = \mathrm{e}^{\mathrm{i}Ht} \Pi({\rm x})\mathrm{e}^{-\mathrm{i}Ht}$. Using integration by parts and neglecting the boundary terms (which can be safely done in the limit $L\to\infty$), we verify the following:
\beq\label{1}
	\dot\Phi(t,{\rm x}):=\Pi({\rm x}),\quad
	\dot\Pi(t,{\rm x}) \stackrel{L\to\infty}= (\nabla_{\rm x}^2-m^2)\Phi(t,{\rm x}).
\eeq
Further, from \eqref{cr}, we find the following equal-time commutation relations at large $L$:
\beq\label{2}
	[\Phi({\rm x}),\Phi({\rm y})] = [\Pi({\rm x}),\Pi({\rm y})]=0,\quad
	[\Phi({\rm x}),\Pi({\rm y})] \stackrel{L\to\infty}= \mathrm{i}\delta^D({\rm x}-{\rm y}).
\eeq
The vacuum $|\vac\ket$ of the $d$-dimensional Klein--Gordon theory, defined in subsection \ref{ssectdens} using mode operators, may be equivalently defined by the conditions
\beq
	\lim_{\tau\to-\infty} \phi(x,-\mathrm{i}\tau)|\vac\ket = 
	\lim_{\tau\to-\infty} \pi(x,-\mathrm{i}\tau)|\vac\ket = 0.
\eeq
Hence we also have
\beq\label{3}
	\lim_{\tau\to-\infty} \Phi(x,-\mathrm{i}\tau)|\vac\ket = 
	\lim_{\tau\to-\infty} \Pi(x,-\mathrm{i}\tau)|\vac\ket = 0.
\eeq
Relations \eqref{1}, \eqref{2} and \eqref{3} (for $L\to\infty$) define the $D$-dimensional Klein--Gordon theory for the canonical fields $\Phi(t,{\rm x})$ and $\Pi(t,{\rm x})$ and the vacuum $|\vac\ket$. That is, the equations \eqref{mapd} for $L\to\infty$ can be seen as defining a homomorphism $\Omega: d\mbox{-dim KG}\to D\mbox{-dim KG}$ from a $d$-dimensional to a $D$-dimensional Klein--Gordon theory. In particular, correlation functions of $\Phi(t,{\rm x})$ and $\Pi(t,{\rm x})$ in the vacuum, seen as integrals of correlation functions in the $d$-dimensional Klein--Gordon theory via \eqref{mapd}, can be calculated using the $D$-dimensional Klein--Gordon theory. 

Note that the facts that the time evolution generated by $H$ is the natural one in the $D$-dimensional theory and that it annihilates $|\vac\ket$, along with \eqref{2}, implies that, on the $D$-dimensional theory (the image of $\Omega$),
\beq\label{HD}
	H|_{{\rm Im}(\Omega)} = \frc12 \int \mathrm{d}^D{\rm x}\,:\big(\Pi({\rm x})^2+(\nabla\Phi({\rm x}))^2 \big):\;.
\eeq
Recall \eqref{HLR}. The local equations of motion generated by $H_{\rl,\rr}$ take the same form on the fields $\Phi(t,{\rm x})$ and $\Pi(t,{\rm x})$ for any $D$, by similar arguments as those above. This implies that we also have
\beq\label{HLRD}
	H_{\rm L,R}|_{{\rm Im}(\Omega)} = \frc12 \int_{x_1\lessgtr 0} \mathrm{d}^D {\rm x}
	\,:\big(\Pi({\rm x})^2+(\nabla\Phi({\rm x}))^2 + m^2\Phi({\rm x})^2\big):\;.
\eeq
Hence, the initial density matrix $\rho_0$ \eqref{rho0} has the same form for any $D$, in terms of the integrated fields. This means that, from the viewpoint of the integrated fields, the $d$-dimensional quench problem is equivalent to the $D$-dimensional quench problem for the Klein--Gordon theory. Hence in particular the steady-state density matrix, on ${\rm Im}(\Omega)$, also has the same form \eqref{rhoness} in terms of the modes of the $D$-dimensional Klein--Gordon fields $\Phi({\rm x})$ and $\Pi({\rm x})$.

Naturally, the normal-ordering operation of the $d$-dimensional theory is mapped under $\Omega$ to the $D$-dimensional normal-ordering. However, one can consider a different set of operators involving products of fields, which does not lie in ${\rm Im}(\Omega)$. Consider the set of all fields of the form
\beq
	L^{\frc{(n-2)(D-d)}2} \int_{-\frc L2}^{\frc L2}
	\mathrm{d}^{d-D}\t x\,
	\Or_1(t_1,{\rm x}_1,\t x)\cdots
	\Or_n(t_n,{\rm x}_n,\t x)
\eeq
for $\Or_i\in\{\phi,\pi\}$ and $n$ a positive integer. This set of fields generates the same {\em Lie algebra} (under commutation) as does the set
\beq
	\Or_1(t_1,{\rm x}_1)\cdots
	\Or_n(t_n,{\rm x}_n)
\eeq
for $\Or_i\in\{\Phi,\Pi\}$ and $n$ a positive integer, within the $D$-dimensional Klein--Gordon theory. With $n=1$, the operators do lie in ${\rm Im}(\Omega)$ and this is just a consequence of the fact that $\Omega$  is a homomorphism of Klein--Gordon theories. Considering commutators between $n=2$ and $n=1$ operators, and the fact that the normal-ordering of a $n=2$ operator only adds a term proportional to the identity operator, we find that equations \eqref{HD} and \eqref{HLRD} indeed agree with the above Lie algebra statement.

This technique of dimensional reduction can be applied straightforwardly to any free-field theory. We will discuss similar ideas in general QFT in a forthcoming work.

Finally, we note that applying the results of this discussion to the case $D=1$, one can use CFT, where the fields separate into right and left movers.

\section{Semiclassical regime from the quantum computation} \label{appsemi}

We may justify this semiclassical logic
to arrive at the expectation value for the current in the steady state
Eq. \eqref{semclass} by explicitly evaluating the
analogue of Eq. (\ref{itir}) with fixed initial boundary condition
$\phi(t=0,x^1=0)=0$ (for simplicity).  In particular, let us focus on
a single momentum mode -- i.e., the solution to the equation of motion
for a single momentum mode: $\phi(x,0) = 2\sin(p^1x)\Theta(x^1)$.  The
dynamics of this problem is completely one dimensional, and so we need
only evolve this forward according to the one dimensional
Klein--Gordon equation, where the effective mass is simply
$\tilde{p}^2+m^2 \equiv M^2$.  For simplicity, for the remainder of
this section we drop the 1 superscripts on $x$ and $p$.

The integral that we need to evaluate is the function which serves as the coefficient of $B_p$ in Eq. (\ref{itir}): \begin{equation}
\mathcal{B}_p \equiv \int\limits_{-\infty}^\infty \frac{\mathrm{dq}}{2\pi} \left(\frac{1}{\mathrm{q}-p-\mathrm{i}\mathbf{0}} - \frac{1}{\mathrm{q}+p-\mathrm{i}\mathbf{0}}\right) \mathrm{e}^{\mathrm{iq}x} \left(a^+_{\mathrm{q},p}\mathrm{e}^{-\mathrm{i}E_{\mathrm{q}}t} + a^-_{\mathrm{q},p}\mathrm{e}^{\mathrm{i}E_{\mathrm{q}}t}\right) 
\end{equation}Let us focus on the limit where $x,t\rightarrow \infty$.  In this case, the exponentials oscillate extremely rapidly and we therefore may employ standard tricks to  bound such integrals.   In particular, let us break the integral over q into many small regions.   Away from $\mathrm{q}=\pm p$, the pole factors and $a^\pm_{\mathrm{q},p}$ are slowly varying and may be approximated well by a constant in a neighborhood of $\mathrm{q}=\mathrm{q}_0$ of width $2\delta \mathrm{q}$.   The contribution to $\mathcal{B}_p$ from this region, from the $a^+$ term, is given by \begin{equation}
\int\limits_{- \delta\mathrm{q}}^{+ \delta\mathrm{q}} \frac{\mathrm{d}\delta\mathrm{q}}{2\pi} \frac{2p a^+_{\mathrm{q}_0,p}}{q^2-p^2} \exp\left[\mathrm{i}\left(\left(x-E^\prime_{\mathrm{q}} t\right)\delta\mathrm{q} - \frac{t}{2} E^{\prime\prime}_{\mathrm{q}_0} \delta\mathrm{q}^2 + \cdots \right)\right]  \label{eqapproxintegral}
\end{equation}A nearly identical answer holds for the contribution due to $a^-$, but with $t\rightarrow -t$.   Primes on $E_{\mathrm{q}}$ stand for derivatives.   If $x - E^\prime_{\mathrm{q}}t \ne 0$, and the ratio $x/t$ is held fixed while $t\rightarrow \infty$, then this integral is bounded from above by \begin{equation}
\frac{4p a^+_{\mathrm{q}_0,p}\delta \mathrm{q}_0}{\pi \left(\mathrm{q}_0^2-p^2\right)} \frac{1}{x-E^\prime_{\mathrm{q}}t}
\end{equation}So long as we avoid the poles, and any point where $x-E^\prime_{\mathrm{q}}t = 0$, then summing over the boxes at each $q_0$ gives us a finite integral that is strictly bounded by $1/t$ and thus vanishes in the limit $t\rightarrow \infty$.   

There are two loopholes to the above argument.  Let us first focus on a point away from the poles where $x=E^\prime_{\mathrm{q}}t$ (note this will always occur only for $|x|\le t$).   In this case, we can instead bound Eq. (\ref{eqapproxintegral}) by the saddle point method, and the integral will decay as $t^{-1/2}$, which again vanishes as $t\rightarrow \infty$.   Evidently, the regions near $\mathrm{q}=\pm p$ are the only dominant contributions to this integral as $t\rightarrow \infty$, with $x/t$ fixed.   In this case, we can thus (after some re-shifting of the variable q) approximate the $a^+$ contributions to $\mathcal{B}_p$ (away from points where $x = \pm E^\prime_p t$): \begin{align}
\mathcal{B}_p &\approx \int\limits_{-\infty}^\infty \frac{\mathrm{dq}}{2\pi (\mathrm{q} - \mathrm{i}\mathbf{0})} \left(\mathrm{e}^{\mathrm{i}(px-E_p t)+\mathrm{iq}(x -E^\prime_pt)} - \mathrm{e}^{\mathrm{i}(-px-E_p t)+\mathrm{iq}(x +E^\prime_pt)} \right) \notag \\
&= \mathrm{e}^{\mathrm{i}(px-E_pt)}\Theta(x-E^\prime_pt) + \mathrm{e}^{\mathrm{i}(-px-E_pt)} \Theta(x+E^\prime_p t).
\end{align}The $a^-$ contributions to $\mathcal{B}_p$ vanish, as $a^-_{p,p}=0$.   The above equation shows that for any ratio of $x/t$ fixed (and not equal exactly to the group velocity of momentum mode $p$), at late times momentum mode $p$ is approximately described by its semiclassical dynamics:  a pair of waves traveling to the left or right at the classical (group) velocity.   This demonstrates that our evaluation of the expectation values of $\langle T^{\mu\nu}\rangle$ is asymptotically exact as $x^1,t\rightarrow \infty$.

\section{Free Dirac fermion results} \label{appfermion}

We may straightforwardly generalize the analysis of this paper to the
case of fermionic models, such as the free massive Dirac fermions. In
this case, there is a $\mathrm{U}(1)$ charge, and we may thermalize
the reservoirs with temperatures $T_\rl,\;T_\rr$ and chemical
potentials $\mu_\rl,\;\mu_\rr$. The results for the energy current and
the energy density are \beqa \bra T^{01}\ket_{\rm s} &=&
\frc{\Gamma\lt(\frc{d}2\rt)}{4\pi^{d/2+1}(d-1)!} \sum_{\ep=\pm}
\int_0^\infty \mathrm{d}p\,p^d \,\frc{\sinh\lt(\frc{\beta_{\rm
      R}-\beta_{\rm L}}2 E_p\rt)}{\cosh\lt(\frc{\beta_{\rm
      R}}2(E_p-\ep \mu_{\rm R})\rt)\cosh\lt(\frc{\beta_{\rm
      L}}2(E_p-\ep \mu_{\rm R})\rt)}\n &=& \frc{d\,\Gamma\lt(\frc
  d2\rt)}{2\pi^{d/2+1}}\,\sum_{\ep=\pm}\lt( \zeta_{\frc m
  {T_\rl},\frc{\ep\mu_\rl}{T_\rl}}(d+1)\,T_\rl^{d+1}- \zeta_{\frc m
  {T_\rr},\frc{\ep\mu_\rr}{T_\rr}}(d+1)\,T_\rr^{d+1}\rt)\n \bra
T^{00}\ket_{\rm s} &=&
\frc{\Gamma\lt(\frc{d+1}2\rt)}{4\pi^{(d+1)/2}(d-1)!}
\sum_{\ep=\pm}\int_0^\infty \mathrm{d}p\,p^{d-1}E_p
\,\frc{\cosh\lt(\frc{\beta_{\rm R}-\beta_{\rm L}}2
  E_p\rt)}{\cosh\lt(\frc{\beta_{\rm R}}2(E_p-\ep \mu_{\rm
    R})\rt)\cosh\lt(\frc{\beta_{\rm L}}2(E_p-\ep \mu_{\rm R})\rt)}\n
&=& \frc{d \,\Gamma\lt(\frc{d+1}2\rt)} {2\pi^{(d+1)/2}}\,
\sum_{\ep=\pm}\lt(\t\zeta_{\frc
  m{T_\rl},\frc{\ep\mu_\rl}{T_\rl}}(d+1)\,T_\rl^{d+1}+ \t\zeta_{\frc
  m{T_\rr},\frc{\ep\mu_\rr}{T_\rr}}(d+1)\,T_\rr^{d+1}\rt), \eeqa where
we define the functions \beq\label{zeta} \zeta_{a,b}(d) :=
\frc1{\Gamma(d)} \int_0^\infty \mathrm{d}p\,\frc{p^{d-1}}{
  \mathrm{e}^{\sqrt{p^2+a^2}-b}+1},\quad \t\zeta_{a,b}(d) :=
\frc1{\Gamma(d)} \int_0^\infty
\mathrm{d}p\,\frc{p^{d-2}\sqrt{p^2+a^2}}{
  \mathrm{e}^{\sqrt{p^2+a^2}-b}+1}.  \eeq In the massless limit with
zero chemical potential (corresponding to $a=b=0$) both functions in
Eq.~\eqref{zeta} specialize to $\zeta_{0,0}(d) = \t\zeta_{0,0}(d) =
(1-2^{1-d})\zeta(d)$, and the result agrees with \cite{collura2}. In
particular, setting $d=1$ and $m=0$, and using $\zeta(2) = \pi^2/6$
with $\Gamma(1/2)=\sqrt{\pi}$, we find again that the coefficients of
the powers of temperature all specialize to $\pi/12$ as required for a
1+1 CFT with central charge $c=1$ \cite{doyon1d1,doyon1d2}.

We may also define the SCGF for energy transfer, $F(z)$. Thanks to the results of \cite{doyontrans}, since this is a free model again one may use the extended fluctuation relations, so that \eqref{Fz} still holds. Energy transfer can be interpreted via Poisson processes again, but only at zero chemical potentials $\mu_\rl=\mu_\rr=0$. This is the same situation as that which was found in 1+1-dimensional CFT \cite{doyon1d1,doyon1d2}. Here we find the weight
\beq \omega(q) =
\lt(2^{d-1}\pi^{\frc{d+1}2} \Gamma\lt(\frc{d+1}2\rt)\rt)^{-1}\,
\sum_{n=1}^{[|q|/m]} \frc{(-1)^{n-1}}{n^{d+1}}\lt(q^2-n^2m^2\rt)^{\frc{d-1}2}
\cdot\lt\{\ba{ll} \mathrm{e}^{-\beta_\rl q} & (q>0) \z \mathrm{e}^{\beta_\rr q} & (q<0)
\ea\rt.  \eeq

\end{document}